\documentclass[usenatbib]{mn2e}

\pdfoutput=1

\usepackage{natbib}
\usepackage{graphicx}
\usepackage{apjfonts}
\usepackage{deluxetable}
\usepackage{multicol}
\usepackage{grant_defs}
\usepackage{verbatim}

\usepackage{aastex_hack}

\title[Residual Cooling amid AGN Feedback in Abell 2597]{Residual Cooling and Persistent Star Formation amid \\ AGN Feedback in Abell 2597}

\author[G.~R.~Tremblay et al.]{G.~R.~Tremblay,$^{1,2,3}$ C.~P.~O'Dea,$^{2,4}$ S.~A.~Baum,$^{3,5}$ T.~E.~Clarke,$^{6}$ C.~L.~Sarazin,$^{7}$
\newauthor J.~N.~Bregman,$^{8}$ F.~Combes,$^{9}$ M.~Donahue,$^{10}$ A.~C.~Edge,$^{11}$ A.~C.~Fabian,$^{12}$ G.~J.~Ferland,$^{13}$
\newauthor  B.~R.~McNamara,$^{4,14}$ R.~Mittal,$^{3}$ J.~B.~R.~Oonk$^{15}$ A.~C.~Quillen,$^{16}$ H.~R.~Russell,$^{14}$ 
\newauthor J.~S.~Sanders,$^{12}$ P.~Salom\'{e},$^{9}$ G.~M.~Voit,$^{10}$ R.~J.~Wilman,$^{11}$ and M.~W.~Wise$^{15}$ 
\\\\$^{1}$ European Southern Observatory, 
Karl-Schwarzschild-Str.~2, 85748 Garching bei M\"{u}nchen, Germany; grant.tremblay@eso.org 
\\$^{2}$ Department of Physics, Rochester Institute
of Technology, 84 Lomb Memorial Drive, Rochester, NY 14623, USA
\\$^{3}$ Chester F.~Carlson Center for Imaging Science,  54 Lomb Memorial Drive, Rochester, NY 14623, USA
\\$^{4}$ Harvard-Smithsonian Center for Astrophysics, 60 Garden St., Cambridge, MA 02138, USA
\\$^{5}$ Radcliffe Institute for Advanced Study, 10 Garden St., Cambridge, MA 02138, USA
\\$^{6}$ Naval Research Laboratory Remote Sensing Division, Code 7213 4555 Overlook Ave SW, Washington, DC 20375, USA
\\$^{7}$ Department of Astronomy, University of Virginia, P.O. Box 400325, Charlottesville, VA 22904-4325, USA
\\$^{8}$ University of Michigan, Department of Astronomy, Ann Arbor, MI 48109, USA
\\$^{9}$ Observatoire de Paris, LERMA, CNRS, 61 Av.~de l'Observatoire, 75014 Paris, France
\\$^{10}$ Michigan State University, Physics and Astronomy Dept., East Lansing, MI 48824-2320, USA
\\$^{11}$ Department of Physics, Durham University, Durham, DH1 3LE, UK
\\$^{12}$ Institute of Astronomy, Madingley Rd., Cambridge, CB3 0HA, UK
\\$^{13}$ Department of Physics, University of Kentucky, Lexington, KY 40506, USA
\\$^{14}$ Physics \& Astronomy Dept., Waterloo University, 200 University Ave.~W., Waterloo, ON, N2L, 2G1, Canada
\\$^{15}$ ASTRON, Netherlands Institute for Radio Astronomy, P.O. Box 2, 7990 AA Dwingeloo, The Netherlands
\\$^{16}$ Department of Physics and Astronomy, University of Rochester, Rochester, NY 14627, USA}

\begin{document}

%
% Bibliography and bibfile
\def\aj{AJ}%
          % Astronomical Journal
\def\araa{ARA\&A}%
          % Annual Review of Astron and Astrophys
\def\apj{ApJ}%
          % Astrophysical Journal
\def\apjl{ApJ}%
          % Astrophysical Journal, Letters
\def\apjs{ApJS}%
          % Astrophysical Journal, Supplement
\def\ao{Appl.~Opt.}%
          % Applied Optics
\def\apss{Ap\&SS}%
          % Astrophysics and Space Science
\def\aap{A\&A}%
          % Astronomy and Astrophysics
\def\aapr{A\&A~Rev.}%
          % Astronomy and Astrophysics Reviews
\def\aaps{A\&AS}%
          % Astronomy and Astrophysics, Supplement
\def\azh{AZh}%
          % Astronomicheskii Zhurnal
\def\baas{BAAS}%
          % Bulletin of the AAS
\def\jrasc{JRASC}%
          % Journal of the RAS of Canada
\def\memras{MmRAS}%
          % Memoirs of the RAS
\def\mnras{MNRAS}%
          % Monthly Notices of the RAS
\def\pra{Phys.~Rev.~A}%
          % Physical Review A: General Physics
\def\prb{Phys.~Rev.~B}%
          % Physical Review B: Solid State
\def\prc{Phys.~Rev.~C}%
          % Physical Review C
\def\prd{Phys.~Rev.~D}%
          % Physical Review D
\def\pre{Phys.~Rev.~E}%
          % Physical Review E
\def\prl{Phys.~Rev.~Lett.}%
          % Physical Review Letters
\def\pasp{PASP}%
          % Publications of the ASP
\def\pasj{PASJ}%
          % Publications of the ASJ
\def\qjras{QJRAS}%
          % Quarterly Journal of the RAS
\def\skytel{S\&T}%
          % Sky and Telescope
\def\solphys{Sol.~Phys.}%
          % Solar Physics
\def\sovast{Soviet~Ast.}%
          % Soviet Astronomy
\def\ssr{Space~Sci.~Rev.}%
          % Space Science Reviews
\def\zap{ZAp}%
          % Zeitschrift fuer Astrophysik
\def\nat{Nature}%
          % Nature
\def\iaucirc{IAU~Circ.}%
          % IAU Cirulars
\def\aplett{Astrophys.~Lett.}%
          % Astrophysics Letters
\def\apspr{Astrophys.~Space~Phys.~Res.}%
          % Astrophysics Space Physics Research
\def\bain{Bull.~Astron.~Inst.~Netherlands}%
          % Bulletin Astronomical Institute of the Netherlands
\def\fcp{Fund.~Cosmic~Phys.}%
          % Fundamental Cosmic Physics
\def\gca{Geochim.~Cosmochim.~Acta}%
          % Geochimica Cosmochimica Acta
\def\grl{Geophys.~Res.~Lett.}%
          % Geophysics Research Letters
\def\jcp{J.~Chem.~Phys.}%
          % Journal of Chemical Physics
\def\jgr{J.~Geophys.~Res.}%
          % Journal of Geophysics Research
\def\jqsrt{J.~Quant.~Spec.~Radiat.~Transf.}%
          % Journal of Quantitiative Spectroscopy and Radiative Trasfer
\def\memsai{Mem.~Soc.~Astron.~Italiana}%
          % Mem. Societa Astronomica Italiana
\def\nphysa{Nucl.~Phys.~A}%
          % Nuclear Physics A
\def\physrep{Phys.~Rep.}%
          % Physics Reports
\def\physscr{Phys.~Scr}%
          % Physica Scripta
\def\planss{Planet.~Space~Sci.}%
          % Planetary Space Science
\def\procspie{Proc.~SPIE}%
          % Proceedings of the SPIE

\date{Accepted for Publication in MNRAS, 9 May 2012}

\pagerange{\pageref{00} -- \pageref{00}} \pubyear{2012}

\pagerange{\pageref{firstpage}--\pageref{lastpage}} \pubyear{2012}

\maketitle

\label{firstpage}

\begin{abstract}
New {\it Chandra} X-ray and {\it Herschel} FIR observations enable 
a multiwavelength study of active galactic nucleus (AGN) heating 
and intracluster medium (ICM) cooling in the brightest cluster galaxy (BCG) 
of Abell 2597 ($z=0.0821$). The new {\it Chandra} observations
reveal the central $\lae 30$ kiloparsec X-ray cavity network to be 
more extensive than previously thought, and associated with 
enough enthalpy to theoretically inhibit the inferred classical cooling flow. 
Nevertheless, we present new evidence, consistent with previous results, that a moderately strong residual cooling flow is persisting at 4\%-8\% of the classically predicted rates in a spatially structured manner amid the feedback-driven excavation of the X-ray cavity network. New {\it Herschel} observations are used to estimate warm and cold dust masses, a lower-limit gas-to-dust ratio, and a star formation rate consistent with previous measurements. [\ion{O}{i}] and CO(2-1) line profiles are used to constrain the kinematics 
of the $\sim10^9$ \Msol\ reservoir of cold molecular gas. 
The cooling time profile of the ambient X-ray atmosphere is used 
to map the locations of the observational star formation 
entropy threshold as well as the theoretical thermal instability 
threshold. Both lie just outside the $\lae 30$ kpc central region permeated by X-ray cavities, and star formation as well as ionized and molecular gas lie interior to both. 
The young stars are distributed in an elongated region that is aligned 
with the radio lobes, and their estimated ages are both younger 
and older than the X-ray cavity network, suggesting both jet-triggered
as well as persistent star formation over the current AGN feedback episode. Bright X-ray knots that are coincident with extended Ly$\alpha$ and FUV continuum filaments motivate a discussion of structured
cooling from the ambient hot atmosphere along a projected axis that is perpendicular to X-ray cavity and radio axis. We conclude that the cooling ICM is the dominant contributor of the cold gas reservoir fueling star formation and  AGN activity in the Abell 2597 BCG. 
\end{abstract}


%%%%%%%%%%%%%%%
\section{Introduction}
\label{section:introduction}
%%%%%%%%%%%%%%%

For  a subset  of galaxy  clusters with  sharply peaked  X-ray surface
brightness     profiles,     the     intracluster     medium     (ICM,
e.g.,~\citealt{sarazin86}) can cool  via bremsstrahlung processes from
$>10^7$ K to $\ll 10^4$ K  on timescales much shorter than a Gyr
within a radius of $\sim$100  kpc.  Simple models predict that runaway
entropy loss  by gas within  this radius accompanies  subsonic, nearly
isobaric  compression   by  the  ambient  hot   reservoir,  driving  a
long-lived classical  cooling flow onto the  central brightest cluster
galaxy    (hereafter    BCG,    see    cooling   flow    reviews    by
\citealt{fabian94,peterson06}).  In these ``cool core'' (CC) clusters,
catastrophic  condensation  of  the  ICM  should  drive  extreme  star
formation  rates ($10^2-10^3$  \Msol\ yr\mone)  amid massive  cold gas
reservoirs ($\sim10^{12}$ \Msol) in the BCG, and high resolution X-ray
spectroscopy should  detect bright  coolant lines stemming  from $\lae 10^3$  \Msol\  cascades  of   multiphase  gas  condensing  from  the  hot
atmosphere.  Yet three decades  of searches for these expected cooling
flow mass  sinks returned with  results that were orders  of magnitude
below  predictions, consistent  only  with residual  cooling at  $\sim
1-10$\%           of            the           expected           rates
\citep[e.g.,][]{allen69,deyoung74,peterson78,haynes78,baan78,shostak80,chrisstefi87,mcnamara89,odea94,allen95,odea97,odea98,peres98,mittaz01,tamura01,peterson01,xu02,sakelliou02,edge03,peterson03,bregman06,sanders08}.

Attempts to  reconcile these results with the  high X-ray luminosities
and  long  lifetimes  associated   with  the  CC  phase  have  invoked
non-gravitational  heating  mechanisms  to  inhibit  or  replenish  an
average $\sim 90$\% of ICM  radiative losses over the cluster lifetime
\citep[e.g.,  review by][]{peterson06}.   One promising  candidate for
quenching cooling  flows in the innermost  regions of cool  cores is a
feedback  loop  regulated  by  the mechanical  dissipation  of  active
galactic          nucleus          (AGN)         power          (e.g.,
\citealt{rosner89,baum91,churazov02,birzan04,rafferty06,dunn06,best07,edwards07,mittal09},
reviews by \citealt{mcnamara07,mcnamara12,fabian12}).  The  paradigm is supported by strong
circumstantial  evidence, including  observations of  kiloparsec scale
X-ray cavities  in spatial correlation with  radio emission associated
with                            AGN                           outflows
\citep{boehringer93,fabian00,fabian06,churazov01,mcnamara00,mcnamara01,blanton01,nulsen05,forman05,forman07,birzan04}.
The power  of these radio sources are statistically  correlated with X-ray
luminosity  from within the  cooling radius,  and the  most radio-loud
BCGs  are  ubiquitously  associated  with  the  strongest  cool  cores
\citep[e.g.,][]{burns90,birzan04,rafferty06,mittal09,sun09,hudson10}.

\begin{table*}
\centering
    \caption{A summary of the new and archival observations used (directly or referentially) in this analysis.
This table also applies to the analysis presented in \citealt{tremblay_feedback}. 
    (1) Facility name;
    (2) instrument used for observation;
    (3) configuration of instrument / facility;
    (4) wavelength regime / spectral line observed or filter used;
    (5) exposure time;
    (6) facility-specific observation or program ID;
    (7) date of observation;
    (8) principal investigator (PI) of observation or program;    
    (9) reference to publication where the data were first published --- 1: \citet{mcnamara01}; 2: \citet{oonk11}; 3: \citet{odea04}; 4: \citet{koekemoer99};  5:  \citet{holtzman96}; 6: \citet{donahue00}; 8: \citet{donahue07}; 
9: \citet{edge10phot}; 10: \citet{edge10spec}; 
11: \citet{sarazin95}; 12: \citet{taylor99}; 13: \citet{clarke05}}
\begin{tabular}{ccccccccc}
\hline
Observatory &
Instrument   &
Mode/Config &
Band/Filter/Line &
Exp. Time &
Prog. / Obs. ID  &
Obs. Date &
Program PI & 
Reference \\
(1) & (2) & (3) & (4) & (5) & (6) & (7) & (8) & (9) \\
\hline
\hline
  \multicolumn{9}{c}{{\sc X-ray Observations}}\\
\hline
{\it Chandra}  &  ACIS-S  &  FAINT  &  X-ray ($0.5-7$ keV)   &  39.8 ks   &  922  & 2000 Jul 28  &  McNamara  &   1  \\
  \nodata             & ACIS-S   &  VFAINT &  X-ray ($0.5-7$ keV)   &  52.8 ks   &  6934  &  2006 May 1  &  Clarke    &  New    \\
  \nodata             & ACIS-S   &  VFAINT &  X-ray ($0.5-7$ keV)   &  60.9 ks   &  7329  &  2006 May 4  &  Clarke    &  New    \\
\hline
\multicolumn{9}{c}{{\sc Far Ultraviolet / Optical / Near-Infrared Observations}}\\
\hline
{\it HST}      &  ACS     &  SBC    &  F150LP (FUV)  &   8141 s    &  11131  &  2008 Jul 21   &  Jaffe   &       2  \\
\nodata        &   STIS   &   FUV MAMA & F25SRF2  (Ly$\alpha$) & 1000 s & 8107    &     2000 Jul 27     &  O'Dea         &     3     \\
  \nodata            &  WFPC2   &    \nodata     &  F410M               &   2200 s    &  6717  & 1996 Jul 27    &   O'Dea     &     4     \\
    \nodata           &  WFPC2   &   \nodata      &  F450W               &   2500 s    &  6228  & 1995 May 07     &   Trauger   &        5  \\
    \nodata           &  WFPC2   &   \nodata      &  F702W  ($R$-band)   &   2100 s    &  6228  & 1995 May 07     &   Trauger   &        5  \\
   \nodata           &  NICMOS  &  NIC2   &  F212N               &   12032 s   &  7457  & 1997 Oct 19    &    Donahue  &        6  \\
    \nodata           &  NICMOS  &  NIC2   &  F160W  ($H$-band)   &   384 s     &  7457  & 1997 Dec 03   &    Donahue   &        6  \\
ESO VLT             & SINFONI  &  \nodata      &   $K$-band            &   600 s    & 075.A-0074  & 2005 Jul / Aug &  Jaffe  &      7    \\ 
\hline
\multicolumn{9}{c}{{\sc Mid- / Far-Infrared Observations}}\\
\hline
{\it Spitzer}  &  IRAC    &   Mapping     &  3.6, 4.5, 5.8, 8 $\mu$m  &  3600 s (each) &  3506  &  2005 Nov 24    & Sparks   &   8  \\
   \nodata            &  MIPS    &   \nodata        &  24, 70, 160 $\mu$m           &  2160 s (each) &  3506  &  2005 Jun 18   & Sparks   &   8  \\
{\it Herschel} & PACS     &Photometry &  70, 100, 160 $\mu$m          &  722 s (each)  &  13421871(18-20)      &  2009 Nov 20   &  Edge     & 9   \\
 \nodata             & SPIRE    &Photometry &  250, 350, 500 $\mu$m         &    3336 s      & 1342187329  &  2009 Nov 30   &  Edge     & 9   \\
  \nodata             & PACS     &Spectroscopy& {}[\ion{O}{i}] $\lambda$63.18 $\mu$m  &  6902 s         & 1342187124  & 2009 Nov 20  &  Edge     &  10 \\
  \nodata             &  \nodata        &     \nodata       & {}[\ion{O}{iii}] $\lambda$88.36 $\mu$m  &  7890 s         & 1342188703  &  2009 Dec 30   & Edge   &  10 \\
  \nodata             &  \nodata       &     \nodata       & {}[\ion{N}{ii}] $\lambda$121.9 $\mu$m &  7384 s         & 1342188942  &  2010 Jan 04   & Edge   & 10  \\
  \nodata            &    \nodata      &     \nodata       & {}[\ion{O}{i}{\sc b}] $\lambda$145.52 $\mu$m & 7382 s         & 1342188704  &  2009 Dec 30   & Edge   &  10 \\
  \nodata            &   \nodata      &      \nodata      & {}[\ion{C}{i}] $\lambda$157.74 $\mu$m  & 6227 s         & 1342187125  &  2009 Nov 20   & Edge   & 10 \\ 
  \nodata            &    \nodata      &      \nodata      & {}[\ion{Si}{i}] $\lambda$68.47 $\mu$m    & 11834 s        & 1342210651  &  2010 Dec 01   & Edge    & New \\
\hline
\multicolumn{9}{c}{{\sc Radio Observations}} \\
\hline
NRAO VLA        &     \nodata     & A array &   8.44 GHz               &   15 min          &  AR279      &   1992 Nov 30    & Roettiger  &  11   \\
\nodata     &   \nodata        & A array &   4.99 GHz               &   95 min          &    BT024    &   1996 Dec 7      & Taylor  &  12,13  \\
 \nodata    &    \nodata       & A array &   1.3  GHz               &   323 min          &    BT024    &  1996 Dec 7     & Taylor  &  12,13    \\
 \nodata    &   \nodata        & A array    &   330 MHz          &   180 min          &   AC647     &   2003 Aug  18 & Clarke   &  13  \\ 
  \nodata    &    \nodata       & B array    &   330 MHz          &   138 min          &   AC647     &   2002 Jun  10 & Clarke   &  13  \cr
  \hline
  \end{tabular}
\label{tab:observations}
\end{table*}

Observationally supported simulations  show that the outflowing plasma
can  drive  sound waves  and  subsonically  excavate  cavities in  the
thermal  gas,  which then  buoyantly  rise,
entrain     magnetic     fields      and     colder     ISM     phases,    and   locally
thermalize  enthalpy associated  with their  inflation as  ambient gas
moves            to            refill           their            wakes
\citep[e.g.,][]{begelman01,ruszkowski02,churazov02,reynolds02,fabian03,fabian08,birzan04,robinson04,dursi08,gitti11}.
The phenomenon  is likely episodic at  a rate coupled to  the AGN duty
cycle,
\begin{comment}
($10^7-10^8$ yr, e.g., \citealt{parma99,best05,shabala08}), % this is the lifetime
\end{comment}
and total energy input when summed over the cluster lifetime can range
from  $\sim10^{55}-10^{61}$ ergs.   In  principle, this  is enough  to
replenish  the total  energy budget  of  ICM radiative  losses for  an
average of  the CC cluster population, although the  spatial distribution and thermalisation 
of
this energy  is one of  several important problems that  challenge the
model  (e.g.,  \citealt{mcnamara07,mcnamara12},  and  references  therein).   The
physics coupling  AGN mechanical energy  to ICM entropy  remain poorly
understood,  and  thermal  conduction,  gas  sloshing,  and  dynamical
friction      likely     play      additional      important     roles
\citep[e.g.,][]{sparks89,ruszkowski02,voigt02,brighenti03,elzant04,soker04,voigt04,parrish10,morsony10,zuhone10,blanton11,sparks09,sparks12}.

\subsection{Residual cooling and star formation amid AGN feedback}

Critical details of the heating  and cooling feedback loop are encoded
in the  mass, energy,  and timescale budgets  of the  low temperature,
high  density gas  phases preferentially  found  in CC  BCGs, such  as
filamentary  forbidden  and Balmer  emission  line  nebulae and  $\sim 10^9-10^{10}$   \Msol\  reservoirs  of   both  vibrationally   excited  and
cold\footnote{Throughout this  paper, we will use  ``hot'' to describe
  $10^7  < T  < 10^8$  K (X-ray  bright) ICM/ISM  phases,  ``warm'' to
  describe $10^4 \lae T <  10^5$ K (optical and UV bright) components,
  and ``cold''  to describe  $10\lae T \lae  10^4$ K  (N/M/FIR bright)
  components.  We  will not
  discuss  the  critically important  $10^5-10^7$  K  regime at  great
  length     in     this     paper.}      molecular     gas     (e.g.,
\citealt{hu85,baum87,heckman89,jaffe97,donahue00,edge01,edge02,jaffe01,wilman02,mcnamara04,jaffe05,rafferty06,egami06,edge10spec,edge10phot,oonk10,mittal11,wilman09,wilman11,salome06,salome11,lim12}).

While residual ICM condensation can contribute a major fraction of the
mass          budget           for          these          phenomena
\citep[e.g.,][]{baum87,heckman89,cavagnolo08,rafferty08,quillen08,odea08,hudson10,mcdonald10,mcdonald11a,mcdonald11b},
their temperatures  and ionization states are  often inconsistent with
the $10 \lae T \lae 10^4$  K phases of a purely radiative cooling flow
\citep[e.g.,][]{donahue91,voit97}.  Thermal  conduction and suprathermal electron
heating        now        appears        to        be        important
(\citealt{ferland08,ferland09,donahue11,fabian11,oonkthesis,mittal11,sparks09,sparks12}; Johnstone et al.~2012, in press),
and  a  critical  role  is   played  by  the  clumpy  and  filamentary
distributions of star formation  ongoing amid these cold reservoirs on
$\lae                  30$                  kpc                 scales
\citep[e.g.,][]{johnstone87,romanishin87,mcnamara89,hu92,crawford93,mcnamara04,hansen95,allen95,voit97,smith97,cardiel98,hutchings00,oegerle01,mittaz01,odea01,odea04,mcnamara04,hicks05,rafferty06,bildfell08,rafferty08,odea08,voit08,loubser09,odea10,oonk11,mcdonald10,mcdonald11a}.

Estimated  star formation  rates range  from a  few to  tens  of solar
masses  per  year,  and   strongly  correlate  with  upper  limits  on
spectroscopically  derived  ICM  mass  deposition rates,  as  well  as
CO-inferred  molecular gas  masses \citep[e.g.,][]{odea08}.   This
suggests  a direct  causal  connection   between  
reduced  cooling  flows and  star
formation, as BCGs with young  stellar populations are always found in
cool   core  clusters   \citep{bildfell08,loubser09}.    New  evidence
suggests  that  star  formation  occurs  when  multiphase  clouds  and
filaments  precipitate out  of the  ICM as  its central  entropy drops
below    a   critical   threshold    ($\lae   20-30$    keV   cm\mtwo,
\citealt{voit08,cavagnolo08,rafferty08,sharma11,gaspari11}).

But while the pathway of entropy  loss from hot ambient medium to cold
star forming clouds may be  strongly influenced by AGN feedback, it is
not known whether residual cooling persists at constant low levels, or
in  elevated episodes  anti-correlated to  the AGN  duty  cycle (e.g.,
\citealt{odea10,tremblay11}).   Moreover,  models invoking  radio-mode
feedback to quench cooling flow signatures such as star formation must
be reconciled with evidence that, in several systems (e.g., Abell 1795
and Abell  2597, \citealt{odea04}),  the propagating radio  source may
trigger  compact, short-duration  starbursts as  it propagates  amid a
dense                           medium                          (e.g.,
\citealt{elmegreen78,voit88,deyoung89,mcnamara93,odea04,batcheldor07,holt08,holt11}).          The
picture is  further complicated by  the role played by  (e.g.) thermal
conduction and  cold accretion scenarios like  gas-rich mergers, whose
importance  relative to  residual hot-mode  ICM cooling  has  driven a
debate    that    is    still    unsettled   after    three    decades
\citep[e.g.][]{holtzman96,sparks89,sparks12}.     Advances   in   understanding
largely  rely upon  multiwavelength data  that sample  all temperature
phases of  the ISM in CC  BCGs, the transport  processes between these
phases, and their associated mass and energy budgets.

\subsection{The brightest cluster galaxy in Abell 2597}

To that end,  in this paper we present a  multiwavelength study of the
archetypal cool core  cluster Abell 2597. New {\it  Chandra} X-ray and
{\it  Herschel}  FIR  observations,  combined  with a  vast  suite  of
archival  data, enable  a  radio-through-X-ray  study of  the
ICM/ISM on the scale of its AGN feedback interaction region (i.e., its
central $\lae  30$ kpc X-ray  cavity network).  Abell  2597 (hereafter
A2597)  is an  Abell \citep{abell58,abell89}  richness class  0 galaxy
cluster with an X-ray surface brightness profile that is sharply peaked 
about its 
%of  total  X-ray  luminosity $L_{\mathrm{X}}  =  6.45  \times
%10^{44}$  ergs  s$^{-1}$  (2-10  keV, \citealt{david93}).   The  X-ray
%surface  brightness  profile is  sharply  peaked  about the  
centrally
dominant  elliptical BCG at  redshift $z=0.0821$  \citep{voit97}.  The
galaxy  is host to  one of  the nearest  known compact  steep spectrum
(CSS, \citealt{chrisreview}) radio sources,  PKS 2322-122, which exhibits a
compact (10 kpc) and bent Fanaroff-Riley class I (FR I) morphology at 8.4 GHz
(\citealt{fanaroff74,wright90,griffith94,odea94,sarazin95}).

A2597 is  one of  the only  CC clusters with  a convincing  {\it FUSE}
detection of  [\ion{O}{vi}]$\lambda1032$ \AA\ emission  stemming from a
$10^5-10^7$  K gas  component \citep{oegerle01},  and  high resolution
{\it XMM-Newton}  X-ray specroscopy reveals  weak \ion{Fe}{xvii} features
amid a soft X-ray excess \citep{morris05}.  It is therefore one of the
best  known candidates  for harboring  a moderately  powerful residual
cooling   flow   with   a    mass   deposition   rate   of   $90\pm15$
\Msol\ yr\mone\ within a 100 kpc cluster-centric radius \citep{morris05}.  The BCG harbors  a $1.8 \pm 0.3
\times10^9$   \Msol\   cold   H$_2$   component,  inferred   from   CO
observations\footnote{Updated  CO-inferred molecular hydrogen  mass is
  from P.~Salom\'{e} (2012, pvt.~comm.),  using unpublished
  IRAM observations. We discuss this in Section 4.} (e.g.,  \citealt{edge01,salome03}), as well as a
young stellar component cospatial with an H$\alpha$-bright filamentary
emission                          line                          nebula
\citep[e.g.][]{heckman89,voit97,koekemoer99,odea04,oonk11}.       These
features   reside  amid   a  network   of  prominent   X-ray  cavities
(\citealt{mcnamara01,clarke05}), making A2597 an ideal subject for studies
of AGN/ISM interactions \citep{tremblay_feedback}. For these  reasons and others, A2597 enjoys a
long history  of cross-spectrum analysis in the  literature (X-ray ---
\citealt{crawford89,sarazin95,sarazin97,mcnamara01,clarke05,morris05};
UV/optical                                                          ---
\citealt{mcnamara93,deyoung95,voit97,cardiel98,mcnamara99,koekemoer99,oegerle01,odea04,jaffe05,oonk10,oonk11};
IR                                                                  ---
\citealt{mcnamara93,voit97,mcnamara99,koekemoer99,odea04,jaffe05,donahue07,donahue11};
sub-mm ---  \citealt{edge01,salome03,edge10spec,edge10phot}; radio ---
\citealt{odea94HI,sarazin95,taylor99,pollack05,clarke05}).

In  Section  2  we  describe  the new  and  archival  multiwavelength
observations used in this paper. In Section 3 
we use the new {\it Chandra} X-ray observations to assemble an AGN 
feedback energy budget by using the kpc-scale cavity network 
as a lower-limit calorimeter to the AGN kinetic energy input. In Section 4
we present new {\it Herschel} FIR data and use it to place constraints 
on the residual cooling flow. These results are used to frame a discussion 
on star formation, which we present in Section 5. A concluding discussion 
and summary are provided in Sections 6 and 7.  
Throughout this  work, we adopt $H_0 = 71~h_{71}^{-1}$
km  s$^{-1}$ Mpc$^{-1}$,  $\Omega_M =  0.27$, and  $\Omega_{\Lambda} =
0.73$.   In  this  cosmology,  1\arcsec\  corresponds  to  $\sim  1.5$
kpc  at  the  redshift  of  the A2597  BCG  ($z=0.0821$).  This
redshift corresponds  to an angular size  distance of $D_A\approx315$ Mpc
and a luminosity distance of $D_L\approx369$ Mpc.

\begin{figure*}
\begin{center}
\includegraphics[scale=0.50]{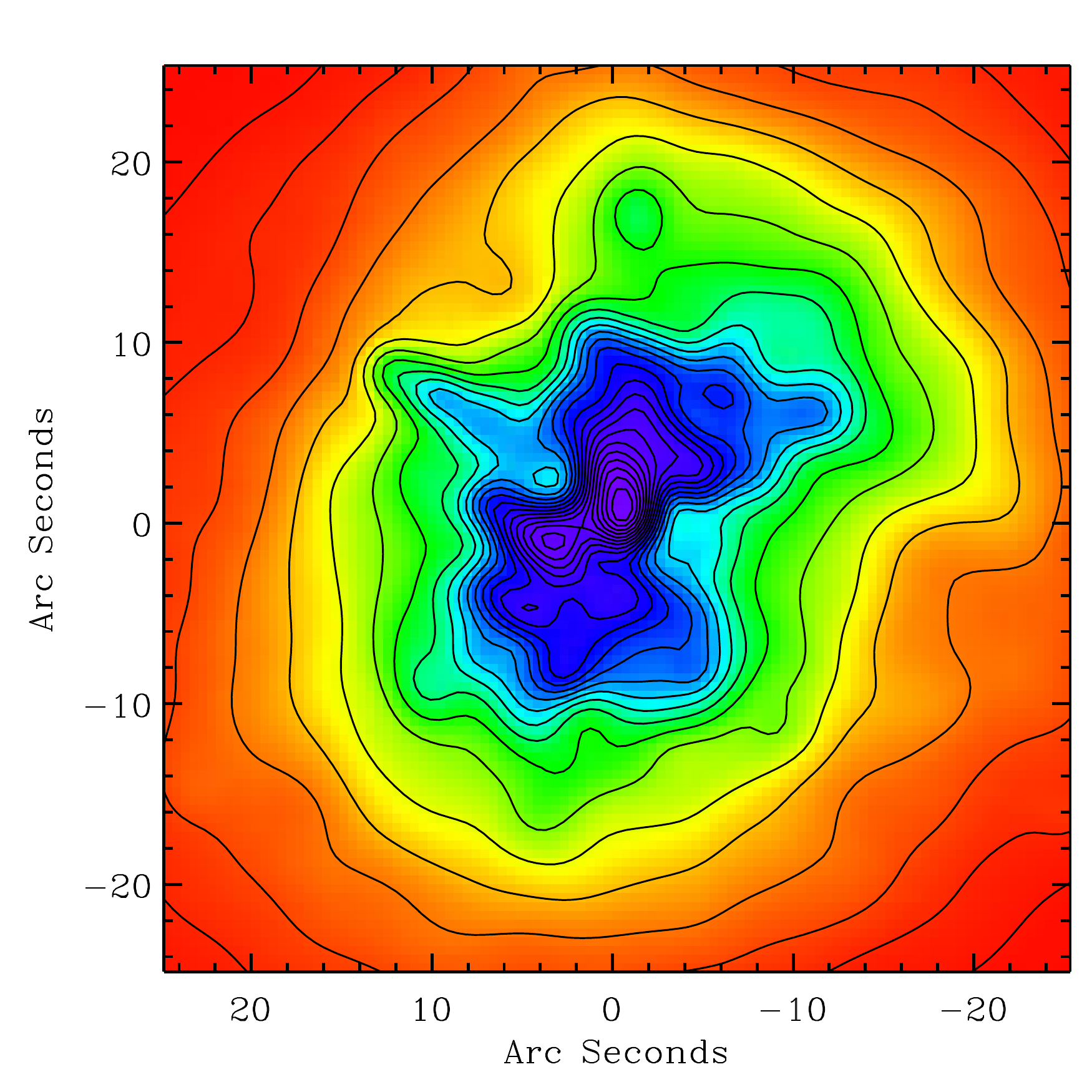}\hspace*{-3mm}
\includegraphics[scale=0.50]{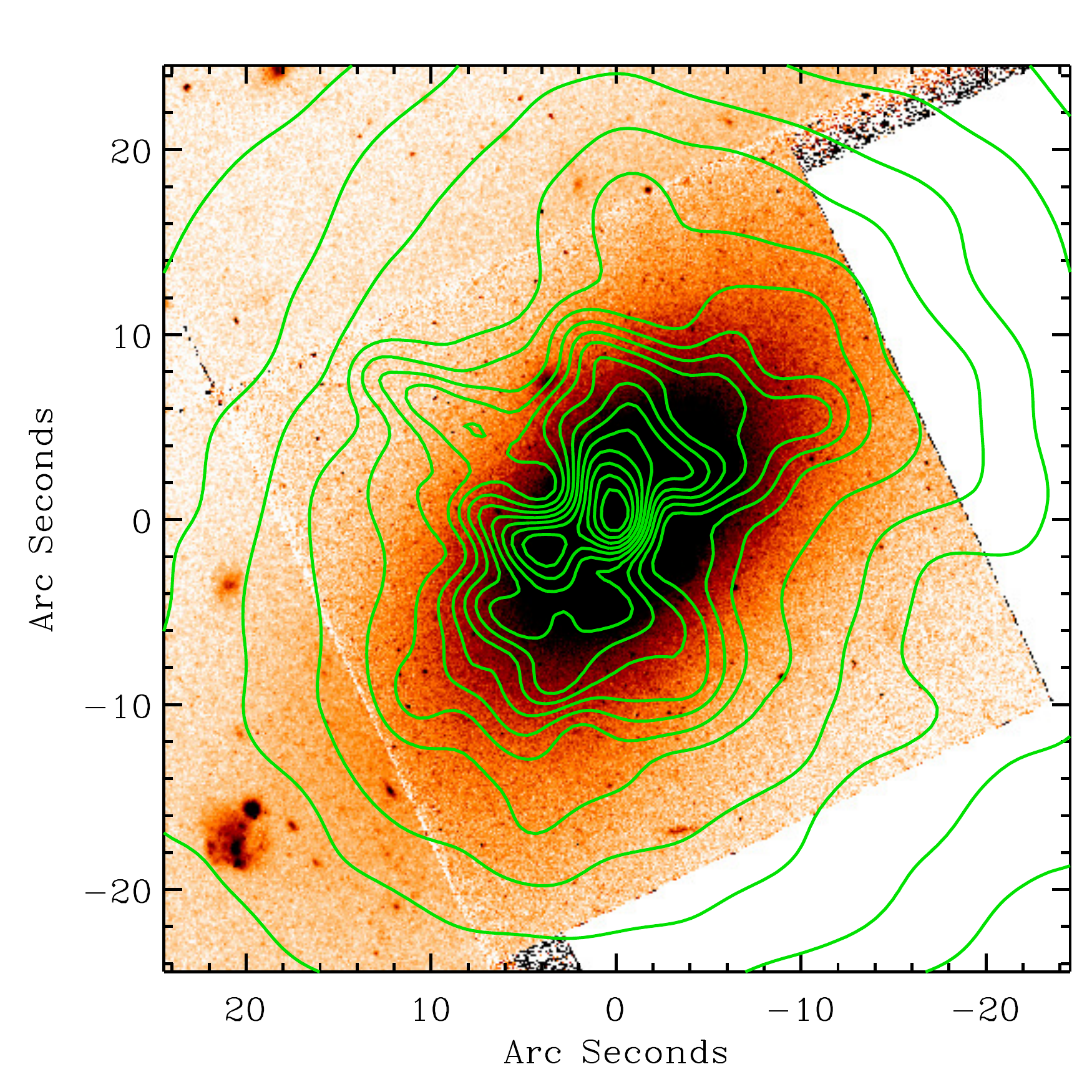}\vspace*{-7mm}
\includegraphics[scale=0.50]{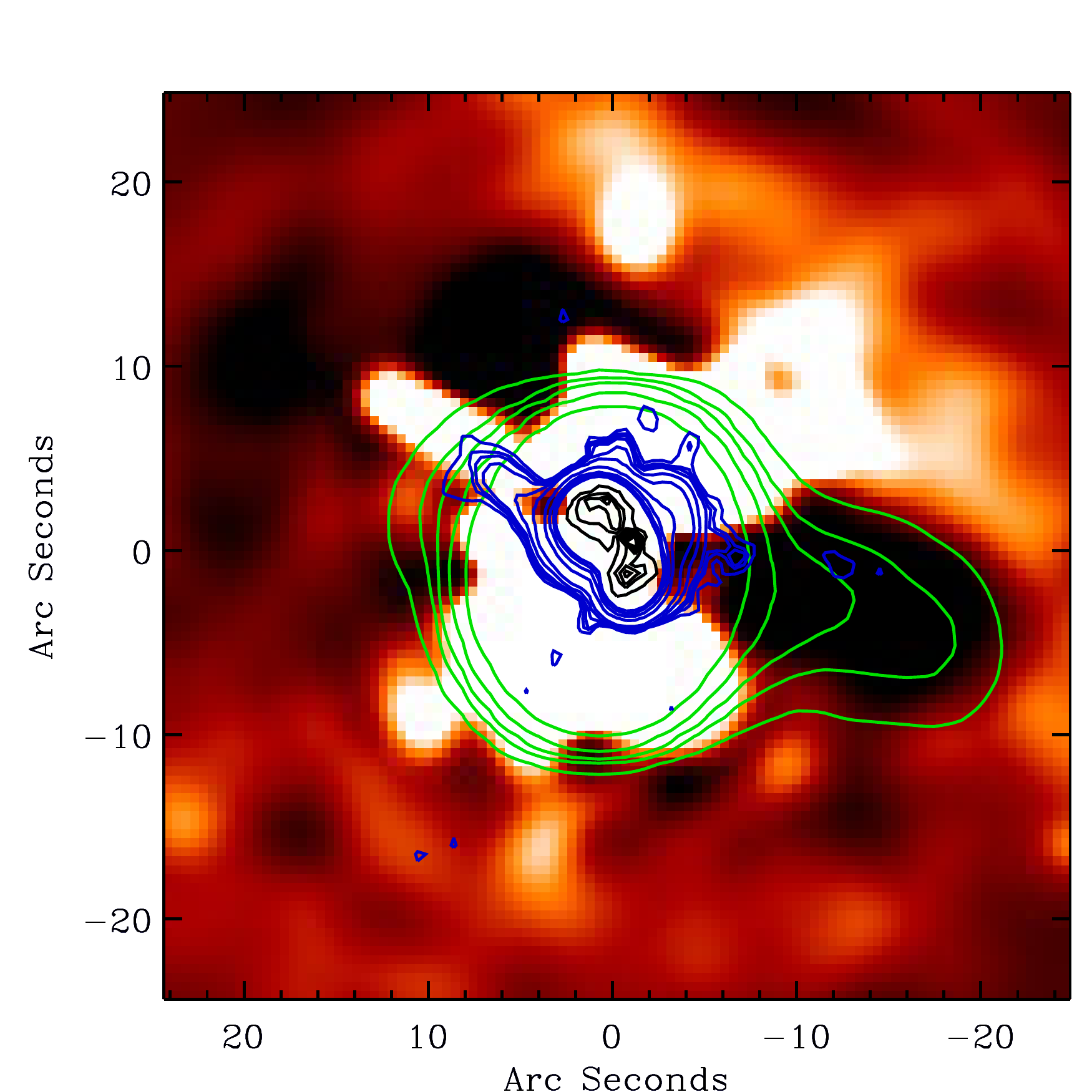}\hspace*{-3mm}
\includegraphics[scale=0.50]{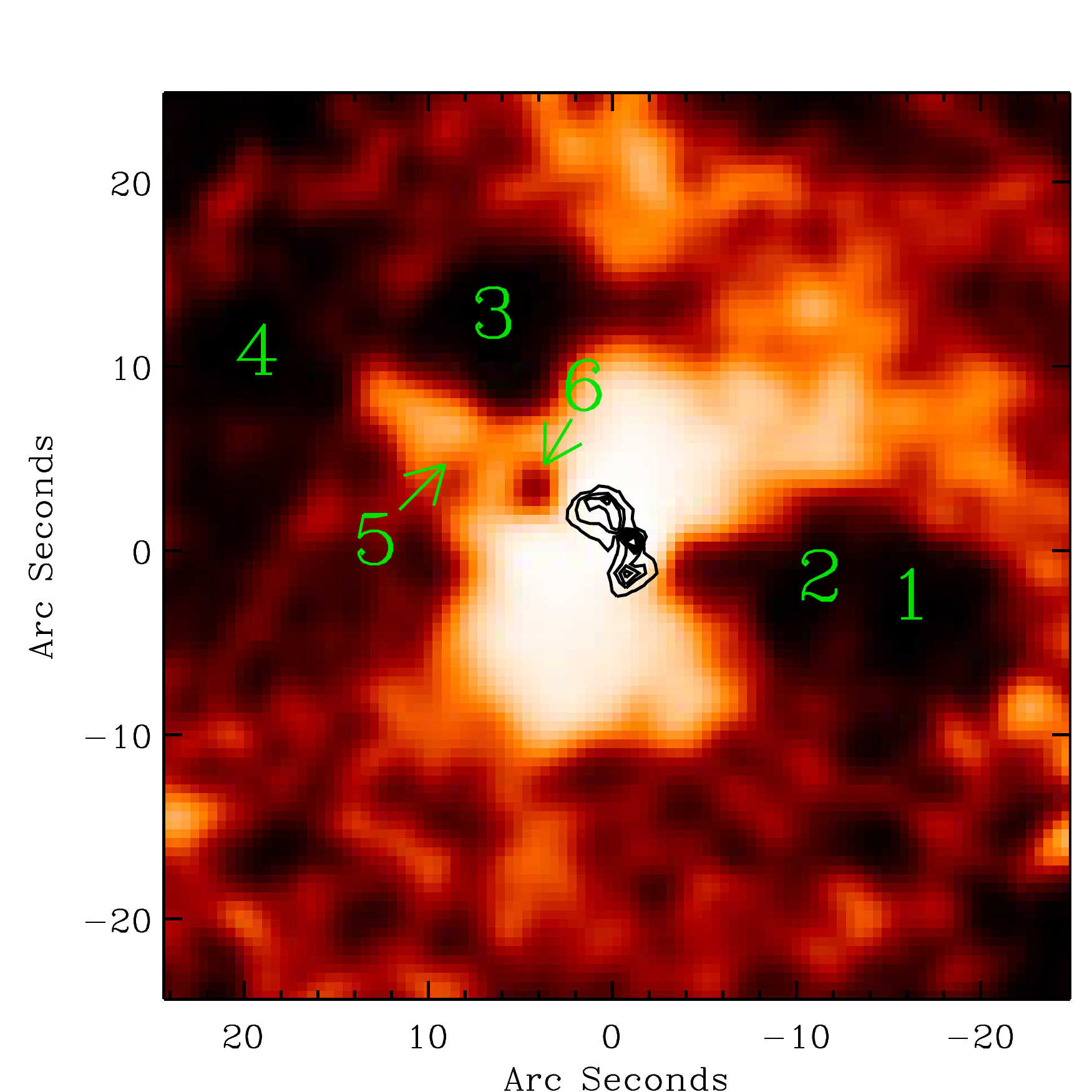}
\end{center}
\caption{({\it top  left}) Exposure-corrected  0.5-7 keV image  of the
  three       merged      {\it      Chandra}   observations (see Table 1).  
The data  have been  smoothed with  an adaptive
  gaussian  kernel.  Black contours  are overlaid  to better  show the
  spatially anisotropic nature of  the emission. The innermost contour
  marks      a       flux      of      $4.6\times10^{-7}$      photons
  sec\mone\ cm\mtwo\ pixel\mone, and  the contours move outward with a
  flux       decrement       of       $2.0\times10^{-8}$       photons
  sec\mone\ cm\mtwo\ pixel\mone.  ({\it top right}) The same 0.5-7 keV
  X-ray  contours   (in  green),  overlaid  on   the  broad-band  {\it
    HST}/WFPC2 F702W  ($R$-band) exposure of the  Abell 2597 brightest
  cluster galaxy  (shown in orange).  Note that the anisotropy  of the
  X-ray  emission is  largely confined  to  the inner  regions of  the
  galaxy.  ({\it bottom left}) Unsharp mask
  of  the X-ray  data, made  by subtraction  of a  20\arcsec\ Gaussian
  smoothed version  from the adaptively smoothed version.  330 MHz VLA
  contours  have been  overlaid  in  green, while  the  1.3 GHz  radio
  contours are plotted in blue,  and the 8.4 GHz radio contours appear
  in black.  ({\it bottom right}) Unsharp mask made by subtracting the
  same  20\arcsec\  Gaussian smoothed  version  of  the  image from  a
  5\arcsec\ Gaussian smoothed version, then dividing by the sum of the
  two  images.  The  major ($10\sigma$ excess or deficit)  morphological  features which
  will be  the subject of further  spatial analysis in  this paper have been  labeled 1-6.  The color  schemes of the two bottom  panels show regions of
  X-ray  surface  brightness  excess  in  red/orange,  while  deficits
  (cavities) appear in black.    All panels share  an identical field
  of view, centered at RA=23h  25m 19.75s, Dec =-12$^\circ$ 07' 26.9''
  (J2000). }
\label{fig:chandra}
\end{figure*}

\section{Observations \& Data Reduction}
\label{section:observations}

%In this  paper, we draw  upon a wide  array of archival data  that has
%been previously presented in the works listed above.  In addition, we present
%previously unpublished deeper  {\it Chandra} observations allowing for
%more  detailed  X-ray spatial and spectral  analysis.
% New  {\it  Herschel}  data,
%preliminarily published in \citet{edge10phot,edge10spec}, is presented
%here  in a format  that frames  the associated  results in  the larger
%context of a multiwavelength study.  

Table  \ref{tab:observations}  contains  a  summary  of  all  new  and
archival observations used in this  paper.  We refer the reader to the
publications listed in column (9) for technical information pertaining
to   the  archival   observations.   The   new  {\it   Chandra}  X-ray
observations,  totalling 150  ksec (128  ksec flare-free)  in combined
effective exposure  time, were reduced  using the standard  {\sc CIAO}
v4.2    threads   with    v4.3.1   of    the    calibration   database.  
More  details can be found in  \citet{tremblay_feedback}, which presents the
spatial and spectral analysis of these new observations.

{\it Herschel  Space Observatory} observations of  A2597 were obtained
in November  2009 with the Photodetector Array  Camera \& Spectrometer
(PACS),  as well  as  the Spectral  and  Photometric Imaging  Receiver
(SPIRE).  These observations  were part  of an  Open Time  Key Project
(OTKP) investigating the FIR line and continuum properties of a sample
of 11  BCGs in  well-known X-ray and  optical line selected  cool core
clusters (PI:  A.~Edge, 140 hrs).  Preliminary results  for A2597 have
been  published  in  the  works  submitted  as  part  of  the  science
demonstration  phase  \citep{edge10phot,edge10spec}.   The  data  were
processed with  the {\it Herschel}  Interactive Processing Environment
(HIPE)  package  version  7.1.0.   See  \citet{edge10phot,edge10spec},
\citet{mittal11}, and Oonk  et al.~(2012, in prep) for  details on the
data reduction.  The present paper uses PACS and SPIRE data to fill in
critical gaps in  the FIR SED for A2597,  enabling constraints on dust
masses  and temperatures.  The  [\ion{O}{i}] $\lambda  68.4~\mu$m PACS
observation, because it is at high spectroscopic resolution, is also used to place 
constraints on the cold gas kinematics.

%%%%%%%%%%%%%%%%%%%%%%%%%%%%%%%%%
\section{AGN Feedback Energy Budget}
\label{section:xrayspatial}
%%%%%%%%%%%%%%%%%%%%%%%%%%%%%%%%%

Kiloparsec   scale  X-ray   cavities  can   be  used   as  lower-limit
calorimeters on  the kinetic energy associated with  AGN outflows, and
the    duty     cycle    over    which     these    outbursts    occur
\citep[e.g.,][]{churazov02,birzan04,dunn06,rafferty06,mcnamara07,mcnamara12,fabian12}.    A
study  of   the  original  40   ksec  {\it  Chandra}   observation  by
\citet{mcnamara01}  and  \citet{clarke05}   (hereafter  M01  and  C05,
respectively)  revealed  western and  northeastern  ghost cavities  in
A2597.  The  new 150 ksec {\it  Chandra} data reveals
the  cavity network  to  be more  extensive  than originally  thought,
changing  the interpretation  of the  AGN feedback  energy  budget and
outburst history.  We discuss these two new results in this section.

\subsection{The X-ray cavity network}

In Fig.~\ref{fig:chandra}  we present the combined 150  ksec 0.5-7 keV
X-ray data from \citet{tremblay_feedback}.  All panels are aligned with
an  identical 50\arcsec  $\times$ 50\arcsec\  (75 kpc  $\times$75 kpc)
field  of view  (FOV) centered  on the  X-ray centroid  at  RA=23h 25m
19.75s, Dec  =-12$^\circ$ 07' 26.9''  (J2000). East is left,  north is
up.   The top  left panel  shows the  exposure corrected,  merged data
adaptively smoothed with a variable width gaussian kernel whose radius
self-adjusts  to  match  the   local  event  density.   Black  surface
brightness contours are overlaid to make individual features easier to
view. The innermost contour marks a flux of $4.6\times10^{-7}$ photons
sec\mone\ cm\mtwo\  pixel\mone, and the contours move  outward with a  decrement of
$2.0\times10^{-8}$ photons sec\mone\ cm\mtwo\ pixel\mone.

In the top  right panel of Fig.~\ref{fig:chandra} we  overlay the
adaptively smoothed X-ray contours on the {\it Hubble Space Telescope} ({\it HST}) WFPC2 $R$-band
observation of  the A2597  BCG (shown in  red/orange).  Note  that the
anisotropy of the X-ray emission is  confined to the scale of the BCG,
while  the  outermost  regions  assume  a  smoother,  more  elliptical
shape.  The major  axes of  the X-ray  and BCG  stellar  isophotes are
aligned.

In the  bottom two panels  of Fig.~\ref{fig:chandra} we show  the same
0.5-7 keV  data processed in  two ways: (1)  a highly processed unsharp mask  made by
subtracting a  30\arcsec\ gaussian smoothed image  from the adaptively
smoothed image shown in  Fig.~\ref{fig:chandra}{\it a}, and (2) a more
conventional unsharp  mask made  by subtracting a  20\arcsec\ gaussian
smoothed  image  from  a  5\arcsec\ (non-adaptive)  gaussian  smoothed
version.  The  subtracted data is then  divided by the sum  of the two
images.   Regions   of  X-ray  surface  brightness   excess  over  the
subtracted   smooth  background  appear   in  white,   while  deficits
(cavities) appear in  black.  8.4 GHz, 1.3 GHz, and  330 MHz Very Large Array (VLA) radio
contours from \citet{sarazin95}, \citet{taylor99}, and \citet{clarke05} are overlaid  on  the bottom  left  panel.  
See section 3.2 of \citet{tremblay_feedback} for  a
discussion of  the strong  radio/X-ray  spatial correlations
that are evident in this figure.  We note that these unsharp mask edge
enhancement methods (particularly method  1) are inherently noisy, and
introduce   artifacts  that   complicate   quantitative  morphological
analysis.   Significance of  the observed  features must  therefore be
estimated from the un-processed data.  We show these mostly as viewing
aids for the X-ray cavities discussed in this section. We do note 
that the processed images shown here do not significantly 
differ in apparent morphology from the un-processed raw {\it Chandra} data that 
can be seen in Fig.~2 of \citet{tremblay_feedback}.

In the bottom right panel  of Fig.~\ref{fig:chandra} we label the six
features that are associated  with $\gae10\sigma$ deficits or excesses
relative to  the local  mean. The significance  of these  features was
estimated by comparing unsmoothed counts  in like sized regions at the
same  cluster-centric  radius.   
Throughout this paper we refer to these 
features with a common label and name. These are: (1)  the  ``M01
western ghost  cavity'', (2) the  ``C05 X-ray tunnel'', (3)  the ``M01
northern  ghost cavity'', (4)  the ``eastern  ghost cavity'',  (5) the
``cold filament'',  and (6)  the ``C05  filament base  cavity''. For
clarity, these  names and labels will be  used consistently throughout
this paper. Feature (2), the ``cold filament'', is discussed in Section 5 of \citet{tremblay_feedback} in the context of AGN/ISM interactions and multiphase gas dredge-up by the radio source.

While  (for ``historical''  reasons)  we label  features  (1) and  (2)
independently, our deeper data make it clear that 
western ghost cavity described by M01 and C05 is part of a larger ``teardrop''
shaped  cavity  $\sim 25$  kpc  in  projected  length (C05  originally
suggested this in their discussion of an ``X-ray tunnel''). 
This changes  the interpretation of the AGN outburst
history  (relative to  the conclusions  drawn in  M01), which  we will
discuss below.  So as  to enable  comparison with  past papers
(e.g., M01;  \citealt{birzan04,rafferty06}) that have  treated the M01
cavity as a separate feature, we maintain independent labels (1 and 2)
for the cavity  and tunnel. We stress that these  features are part of
one larger cavity, which we  will label ``1+2'' and call the ``western
large cavity'' for the remainder of this paper.

\begin{table*}
\begin{minipage}{156mm}
\centering
\caption{Spatial properties and energetics of the X-ray cavities. (1) Label of morphological feature that corresponds to that assigned in the lower right panel of Fig.~\ref{fig:chandra};
(2) name given to the corresponding feature; 
(3) projected radial distance (length) of the feature from the radio core to the estimated center (edge) of the feature;
(4) estimated radius of the feature;
(5) estimated work associated with cavity assuming subsonic inflation; 
(6) age of the cavity if it rises at the local sound speed;
(7) buoyant, subsonic cavity rise time;
(8) time needed to refill the displaced cavity volume;
(9) X-ray cavity power, assuming the cavity is filled with relativistic plasma. See Section 3 for more details on these calculations.}
\begin{tabular}{ccccccccc}
\hline
 & 
 & 
$R$ ($D$) &
$r$ & 
$pV$ & 
$t_{c_s}$ & 
$t_{\mathrm{buoy}}$ & 
$t_{\mathrm{refill}}$ & 
$P_\mathrm{cav}$ \\
Label & 
Name &
(kpc) &
(kpc) &
($\times10^{57}$ ergs) & 
($\times10^7$ yr) &
($\times10^7$ yr) &
($\times10^7$ yr) &
($\times10^{42}$ erg s$^{-1}$) \\
(1) & (2) & (3) & (4) & (5) & (6) & (7) & (8) & (9) \\
\hline
\hline
%1& M01 western ``ghost'' cavity & 24 & 4.8 & 3.1 & 2.7 & 8.8  & 6.6  & 6.52 \\
%2& C05 ``X-ray tunnel'' & \nodata  & \nodata & 5.25 & \nodata & \nodata & \nodata  & \nodata    \\
1+2 & Western ``large cavity''    &  9  &  9    &  35.9  &   1.0    &  1.4   & 5.6  &  170.6    \\ 
3& M01 northern ``ghost'' cavity & 21  & 6.6 & 7.0 &  2.7  & 6.1  &  7.3  & 16.5 \\
4& Eastern ``ghost'' cavity  & 35 & 3.6 & 0.79 & 3.8   &  17.8  &  6.9  & 1.05 \\
6& Filament base cavity  & 9  & 2.3 & 0.30  &  1.1  &  2.7  &  2.8  & 1.73  \cr
\hline
\end{tabular}
\end{minipage}
\label{tab:cavityanalysis}
\end{table*}

\subsection{Age dating the X-ray cavities}
\label{section:cavityage}

We   adopt   the   simple   model   used   by   \citet{birzan04}   and
\citet{rafferty06} in  estimating rough  ages for the  X-ray cavities,
assuming  their  density is  very  low  relative  to the  ambient  gas
density. Three timescales can be  considered, the most simple of which
assumes that  the bubble rises  in the plane  of the sky at  the sound
speed $c_s\simeq\sqrt{kT/\mu  m_\mathrm{p}}$, where $\mu=0.62$  is the
mean molecular weight in units of the proton mass, $m_\mathrm{p}$.  If
the bubble is at projected cluster-centric radius $R$, then this sonic
rise timescale is given by
\begin{equation}
t_{c_s} \simeq R/c_s.
\end{equation}
 As  the  initial  stages  of  cavity  inflation  are  thought  to  be
 supersonic, followed  by subsonic buoyant rise,  this simple approach
 may best reflect an average of the two phases.  Alternatively, if the
 supersonic inflation period is  a negligible fraction of the cavity's
 age and drag  forces limit the cavity's terminal  velocity $v_t$, the
 buoyant timescale can be written as
\begin{equation}
t_\mathrm{buoy} \simeq \frac{R}{v_t} \simeq R \sqrt{\frac{AC}{2gV}},
\end{equation} 
where $g$ is  the local gravitational acceleration, $V$  is the cavity
volume,  $A$ is  its cross-sectional  area, and  $C=0.75$ is  the drag
coefficient     adopted    from    \citet{churazov01}     by    (e.g.)
\citet{birzan04}.   Finally,   cavity  ages  can   be  constrained  by
estimating  the time  required to  refill  the volume  displaced by  a
bubble of radius $r$ as it rises a height equal to its diameter,
\begin{equation} 
t_\mathrm{refill} \simeq 2 R \sqrt{\frac{r}{GM\left(<R\right)}} = 2 \sqrt{\frac{r}{g}},
\end{equation}
where $M\left(<R\right)$ is the mass enclosed within a sphere of radius $R$.

\citet{birzan04}  made all  three estimates  for  the NE  and W  ghost
cavities in A2597 (features  1 and 3 in Fig.~\ref{fig:chandra}, bottom
right  panel),   using  the  early  short  (40   ksec)  {\it  Chandra}
observation.  So as  to provide an independent check  of their results
using the deeper X-ray data, we follow their procedure almost exactly,
and repeat their calculations for  the M01 ghost cavities, the western
large  cavity  (features 1  and  2), as  well  as  the newly  detected
cavities  (4) and  (6) (as  labeled in  Fig.~\ref{fig:chandra}).  Like
\citet{birzan04}, we  adopt the A2597 BCG  stellar velocity dispersion
of  $\sigma\approx224\pm19$ km  s$^{-1}$ from  \citet{smith90}  in our
calculation of the local gravitational acceleration $g$:
\begin{equation} 
g \simeq \frac{2\sigma^2}{R}, 
\end{equation} 
assuming that the galaxy is an isothermal sphere. In \citet{tremblay_feedback} we calculate
a local  value for  $g$ by estimating  the total  enclosed gravitating
mass from a $\beta$-model fit  to the X-ray surface profile.  For this
discussion  however, it  is sufficient  to adopt  the \citet{birzan04}
method  for estimating $g$  (the two  methods turn  out to  be roughly
consistent at the radius of  the X-ray cavity network anyway).  Unlike
\citet{birzan04}, we  use the $kT$ inferred from  the cavity positions
on the X-ray temperature map  presented in \citet{tremblay_feedback} to calculate the sound
speed in the X-ray gas.  We use  the 2D temperature map in lieu of the
1D radial temperature  profile  because A2597 is
azimuthally anisotropic  in X-ray temperature (not  to mention surface
brightness) on these  scales.  Furthermore, in calculating the $pV$ work
associated with each  cavity, we use the projected  density profile to
estimate  the pressure  $nkT$ at  the cluster-centric  radius  of each
cavity.

The  results  of  these  calculations are  given in Table 2.  Our findings
are roughly consistent with those of \citet{birzan04}. For example, in
age  dating the M01  western ghost  cavity we  find $t_{c_s}\approx27$
Myr, $t_\mathrm{buoy}\approx88$  Myr, and $t_\mathrm{refill}\approx66$
Myr, while  \citet{birzan04} finds 26,  66, and 86  Myr, respectively.
None of these estimates account for projection effects, and all assume
that the bubble rises purely in the plane of the sky.  This results in
underestimation of cavity  age by generally less than  a factor of two
\citep{birzan04,rafferty06,mcnamara07}.

\subsection{Cavity heating energy reservoir}

\label{section:cavityenergybudget}

In column (9) of Table 2 we calculate the mean 
instantaneous power of each cavity $P_\mathrm{cav}$, 
\begin{equation}
P_{\mathrm{cav}} = \frac{4pV}{\langle t\rangle}
\end{equation}
where $\langle t  \rangle$ is the average of  the cavity ages listed in columns 6, 7,  and 8. We have  assumed the
cavities  are  filled with  a  relativistic  plasma,  such that  their
enthalpy can  be approximated  as $4pV$ (e.g.,
\citealt{mcnamara07}).

The total sum of all cavity thermal energies ($pV$) listed in column (5)
is  $4.4  \times10^{58}$  ergs.
This serves as a rough, lower-limit estimate on the kinetic energy injected by the AGN into
the ambient X-ray gas during the (a) current AGN episode or 
perhaps (b) during the past two or more episodes of activity.
This is close  to the inferred mechanical  energy of the  central 8.4 GHz
radio source, which \citet{sarazin95} estimated to be $9\times10^{57}$
ergs.  In principle, the  roughly estimated mechanical input  from the  radio source
could be capable  of accounting  for the  energy budget
inferred  from the X-ray  cavities, ignoring  timescale arguments.
Whether or not these cavities have been produced by one or more 
AGN episodes is the subject of the following section.

\begin{comment}
 The estimated
instantaneous  power of  the  western large  cavity  (feature 1+2)  is
$1.7\times10^{44}$ ergs s\mone, which is of the same order as the X-ray luminosity
of  A2597 and  comparable  to  similar estimates  made  for the  X-ray
cavities in Hydra  A \citep{wise07}. In the context  of that cool core
cluster,  this  power has  been  shown  to  be capable  of  offsetting
radiative losses  associated with a cooling flow  with mass deposition
rates  exceeding $>100$  \Msol\ yr\mone\  \citep{david01,wise07}. 
\end{comment}

\subsection{The AGN duty cycle and heating timescale budget}

\label{section:timescalebudget}

A network of multiple cavities found at varying cluster-centric radii (as we find in A2597) 
may be produced with an episodically varying AGN, transitioning between 
either ``on'' and ``off'' or high and low modes. 
Cycling times between  the triggering of radio activity,  the onset of
quiescence, and  the subsequent re-ignition of  activity are typically estimated
to   be  on   the  order   of   $10^7-10^8$  yr   in  general   (e.g.,
\citealt{parma99,best05,shabala08,tremblay10}),  and   synchrotron  losses  limit
radio  source lifetimes  to  $\sim  10^8$ yr,  unless  there has  been
re-acceleration of the electron  population. 
Alternatively, a steady-state AGN with a duty cycle near 100\% can also produce a series of discrete cavities (rather like a fish tank aerator or a dripping faucet, e.g., \citealt{peterson06,mcnamara07}). 
In this section we attempt to distinguish between these two possibilities.

Based  on synchrotron loss  timescales,  \citet{clarke05} estimated
the minimum energy, lower-limit lifetime of  the 330 MHz source ( green contours in the bottom left panel of Fig.~\ref{fig:chandra})
to be in the range of 
 $t_{\mathrm{330~MHz}}
\gae 8  \times 10^6$ yr
(if the spectral index is steep down to 10 MHz) 
to  $t_{\mathrm{330~MHz}}
\gae 5  \times 10^7$ yr
(if the spectral index flattens beyond 330 MHz). 
This model assumes equal cosmic ray and magnetic field energy densities (equipartition), a steep spectral index of -2.7 between 1.3 GHz 
and 330 MHz, and that the source is a uniform prolate cylinder 
with a filling factor of unity. This model yields a minimum energy
magnetic field strength of $29 ~\mu$G and a non-thermal 
pressure of $5\times10^{-11}$ dyn cm$^{-2}$. 
We note that equipartition may be a less-than-ideal assumption 
in this case (e.g., \citealt{fabian02}). Furthermore, 
deeper multiband radio 
observations are required to better assess the spectral index 
of the source across the western large cavity. 

The lower end of the 330 MHz lifetime range
quoted above ($t_{\mathrm{330~MHz}}
\gae 8  \times 10^6$ yr) is significantly shorter than 
the estimated buoyant rise time of the western large cavity it fills (see Table 2, column 7). If we assume that this is the actual 
age of the source, and further assume that the estimated 
age of the western large cavity is roughly correct, then we require (a) {\it in situ} 
re-acceleration of electrons in the cavity or (b) a new episode 
of activity to feed the plasma into an already pre-established cavity (which, if empty, should collapse on a sound crossing time $\lae 10^7$ yr). 
As we will elaborate upon below, there is little supporting evidence 
favoring either of these scenarios.

A ``radio source younger than the cavity'' scenario could also be possible 
if adiabatic and/or inverse Compton  losses dominate
over the synchrotron processes, which would result in the 
radio source fading on a timescale shorter than the synchrotron 
lifetime. The importance of these processes relative 
to synchrotron losses depends on the 
$B$ field strength and unknown
details of  how the radio source  fills the bubble. 
Inverse Compton losses become more efficient than 
synchrotron losses when the $B$ field is extremely weak ($B<1~\mu$G, e.g., \citealt{fabian02}), 
and adiabatic losses are not likely a limiting factor 
because the western large cavity is likely rising at less than the sound  
speed (note the absence of any detected fast shocks in the X-ray data, e.g., \citealt{tremblay_feedback}). We therefore consider this 
scenario unlikely.

The upper end of the minimum-energy synchrotron lifetime is 
$t_{\mathrm{330~MHz}} \gae 5  \times 10^7$ yr (which itself is merely a lower limit). The 330 MHz radio source could therefore easily be roughly 
the same age as the cavity it fills. This is the simplest 
scenario, as it is assumed that the propagation of the radio source amid 
the X-ray gas is the mechanism that excavates the cavity. 
Even if the equipartition and spectral index assumptions mean 
that this estimated lifetime is wrong, pressure equilibrium between the radio source and the ambient hot gas 
can be assumed to estimate a very similar $\sim10^7$ yr minimum age (e.g., \citealt{fabian02}). Considering the high uncertainties 
associated with age dating both cavities and radio sources, 
(not to mention projection effects), we suggest 
that the western large cavity was created by the current, ongoing 
episode of AGN activity.

Considering the above, we cannot rule out the possibility that the AGN in A2597 is on nearly $\sim$100\% of the time. The oldest cavity (i.e., feature 4, with an estimated age of $1.8\times10^8$ yr) may indeed be associated with a previous epoch of AGN activity, but it could also be a detached 
bubble excavated by the current episode (the available data does 
not permit us to discriminate between these possibilities).
We note that the position angle of the 8.4 GHz radio lobes are significantly offset from the western large cavity, 330 MHz and 1.3 GHz radio source (Fig.~\ref{fig:chandra}), and  the VLBA small-scale ($\sim 50$ pc) jet axis \citep{taylor99}.  
We find it unlikely that the offset is due to the 8.4 GHz source 
being associated with a previous epoch of activity, considering 
its steep spectrum \citep{odea94HI,clarke05}, estimated minimum-energy age of $>5\times10^6$ yr \citep{sarazin95}, and alignment of the VLBA ($<50$ pc) and 330 MHz 
($>25$ kpc) major axes \citep{tremblay_feedback}. Several past studies have 
suggested that the bend is due to interaction of the radio source 
with the ambient dense gaseous medium, either by deflection 
or back-flow along pre-established pressure gradients (e.g., \citealt{sarazin95,koekemoer99,odea04,clarke05,oonk10}).

\subsection{Black hole accretion rate}

If the AGN is indeed nearly steady state, the central black hole (BH) requires a stable supply of gas from the ambient accretion reservoir. 
Assuming a mass-to-energy conversion efficiency of $\epsilon=0.1$ \citep{wise07}, and assuming 
that the energy associated with the X-ray cavities is provided by the AGN, 
the sum of mean instantaneous cavity powers ($P_{\mathrm{cav}}$) requires a time averaged BH 
accretion rate of 
\begin{equation}
\dot{M}_\mathrm{acc} \sim \frac{P_{\mathrm{cav}}}{\epsilon c^2} \sim 0.003 -  0.03~ \left( \frac{\epsilon}{0.1} \right)^{-1} M_\odot~\mathrm{yr}^{-1}
\end{equation}
where $c$  is the speed  of light.  
Using the $M_{\mathrm{BH}}-\sigma$ relation 
\citep{magorrian98,ferrarese00,gebhardt00}
with the $K$-band host luminosity and stellar 
velocity dispersion yields a rough BH mass estimate of 
$\sim  3\times10^8$ \Msol,  for which the  corresponding Eddington
accretion rate would  be $\sim 10$ \Msol\ yr\mone\ \citep{rafferty06}. 
This suggests that the accretion rate, if
steady, is strongly sub-Eddington, consistent with past results
even for very powerful AGN  (e.g., Hydra A, \citealt{wise07}).
If the AGN is instead strongly variable or episodic, then gas transport
to the BH could be non-steady and the actual accretion rate could vary.

\subsection{There is enough energy to quench a classical cooling flow}

We calculate the classical X-ray derived cooling time for A2597,
\begin{equation}
t_\mathrm{cool} \equiv \frac{5}{2} \frac{nkT}{n^2 \Lambda},  
\label{eq:coolingtime}
\end{equation}
where $n$ and $kT$ are the gas density and temperature profiles obtained by \citet{tremblay_feedback}, 
and  $\Lambda$ is the  $T>0.02$ keV
portion of  the cooling function from  \citet{sutherland93}, using the generalization from \citealt{tozzi01}.

We find that the cooling time at the $\lae  30$ kpc outermost radius  of the X-ray
cavity network is  $\sim 300$ Myr. 
The  instantaneous  cooling luminosity associated  with the uninhibited cooling  flow (with no heating) can  be calculated from
the classical uninhibited mass deposition rate $\dot{M}_\mathrm{cool}$
using
\begin{equation}
L_\mathrm{cool} = \frac{5}{2} \frac{\dot{M}_\mathrm{cool}}{\mu m_\mathrm{p}} kT_\mathrm{vir},
\end{equation}
where  $T_\mathrm{vir}$  is the  virial  temperature  of the  cluster.
Classical  mass deposition  rates for  A2597 range  from $\sim100-500$
\Msol\  yr\mone\ over a  range of  cluster-centric radius  spanning 
$\sim 30-100$  kpc (e.g., \citealt{allen01,mcnamara01,morris05}).  Taking a cluster
virial temperature  of $kT_\mathrm{vir} =  3.5$ keV, we find  that the
corresponding range of  cooling luminosity is roughly $L_\mathrm{cool}
\approx \left(1-4\right)  \times 10^{44}$ ergs s\mone.   The rough sum
of     mean    instantaneous     cavity     powers    (see     Section
\ref{section:cavityenergybudget}       and      column       9      of
Table 2)  is  $P_\mathrm{cav}\approx  2  \times
10^{44}$ ergs  s\mone, suggesting that, in principle,  there is enough
enthalpy associated with the current cavity network to offset the bulk
of  radiative losses  and  effectively quench  the  cooling flow.

The same conclusion holds if we estimate the predicted cooling 
flow luminosity directly from the new {\it Chandra} data, using lower-limit temperatures derived from spectral fits  (i.e., the \texttt{MKCFLOW} model $kT_{\mathrm{low}}$ values in Table 1 of \citealt{tremblay_feedback}). 
This is a  rough, order-of-magnitude assumption which neglects
many  important considerations  such  as how  this  enthalpy might  be
dissipated and distributed in the ambient gas. Our finding
that the X-ray cavities are associated with enough energy 
to inhibit the predicted classical cooling flow luminosity is consistent 
with previous results \citep{birzan04,dunn06}.

%The focus should be
%to explicitly compare the cooling flow predicted luminosity (mkcflow
%lower temp. limit left free) minus the reduced L_cool (mkcflow lower
%temp. limit 0.1keV) with the AGN energy input and see if this is
%consistent with SFR and cool gas. 

\section{New Evidence supporting a Residual Cooling Flow Model for A2597}

In the previous section we estimated that the X-ray cavity network 
is associated with enough enthalpy to (in principle) replenish a significant fraction 
of the radiative losses associated with a classical cooling flow. However,
the warm and cold gas phases in the A2597 BCG strongly suggest 
that some cooling from the ambient X-ray atmosphere has managed to persist, 
even amid AGN heating. We discuss this possibility here.

\subsection{How might cooling persist amid AGN feedback?}

As mentioned in Section 1, results  from  {\it  XMM-Newton}  and  {\it  FUSE}  are
consistent with a moderately strong  residual (sometimes called ``reduced'') cooling flow with a mass
deposition rate of  $\sim 30\pm15$ \Msol\ yr\mone at  $\sim 30$ kpc to
$90\pm15$      \Msol\     yr\mone\      at     $\sim      100$     kpc
\citep{oegerle01,morris05}. The cooling  luminosity  associated with
this   residual   cooling   flow  is   $L_{\mathrm{Cool,Resid}}\approx
(0.3-1)\times10^{44}$  ergs  sec\mone.  As  with  15-30\%  of CC  BCGs
(depending  on the  sample, e.g.,  \citealt{salome06}), the  A2597 BCG
harbors a  substantial $1.8 \pm  0.3 \times 10^9$ \Msol\  reservoir of
cold H$_2$ (inferred from CO observations) within its central $30$ kpc
(\citealt{edge01,salome03},   P.~Salom\'{e}   private   communication,
2012).  If  the bulk of the mass  budget for  this cold gas  is supplied  by the
residual cooling flow (rather than  e.g., a merger), then AGN feedback
cannot  establish an  impassable entropy  floor, and  some  cooling to
$<100$ K must be permitted even if there is enough energy in principle
to quench the classical cooling flow.  We consider four possibilities:

\begin{enumerate}

\item Low levels of residual cooling persist even while AGN feedback is 
heating the ambient environment;

\item cooling occurs in a spatially structured manner, away from those regions which 
are being locally heated by (e.g.) buoyant cavities and sound waves;

\item cooling occurs in episodes which correspond to the times 
when the BH is inactive, or transitioning between active and inactive 
states or between high and low modes;

\item the substantial cold gas reservoir in the A2597 BCG stems from something 
other than a cooling flow, i.e.~one or more gas rich mergers.  

\end{enumerate} 
These are not the only possibilities, and several or all of the four scenarios 
listed above could work in tandem.

As discussed  in Section \ref{section:timescalebudget},  
we cannot rule out the possibility that the AGN duty cycle in 
A2597 is close to 100\%. 
Even if this is not the case, and the AGN is episodic over $10^{7}$ yr 
periods, a cavity's enthalpy reservoir should be dissipated as heat in the ISM/ICM on a timescale
of  order  the  cavity  lifetime \citep{churazov02,reynolds02,birzan04,dunn06}. 
The net  effect of this time buffer could be that
ISM/ICM heating  rates are roughly  constant in time  for cavity
heating  models,  even  if  the  AGN itself  is  episodic  over  these
timescales.  If  we assume  that this is  correct, and  further assume
that most cold baryons in  the nucleus cooled from the hot atmosphere,
then  we might  conclude  that  low levels  of  residual cooling  must
persist  at a  roughly constant  rate. This  would be  consistent with
scenario (i)  or (ii)  above. 
In this case, a near-constant
cooling  rate would  funnel  a  steady flow  of gas  to the  BH
accretion reservoir,  fueling a  long-lived rather than  episodic AGN.
This would be consistent with our suggestion that the AGN in A2597 
is approximately steady-state. We would require  inhomogeneous
heating and cooling (e.g., scenario  ii, iii and/or iv) to allow cooling to re-trigger
the radio source.

The  notion that  cooling could  proceed in  a spatially  discrete and
structured manner (scenario ii) is not controversial, and was predicted
even  by the earliest  cooling flow  models (see  e.g., the  review by
\citealt{fabian94}). While long-lived  sound waves could theoretically
heat the outermost regions of  cool cores in a more homogenous manner,
heating by  short-lived X-ray cavity enthalpy  dissipation is expected
to  be spatially confined  to the  scale and  location of  the cavity,
regardless of the Reynolds number  of the plasma (see e.g.,~the review
by \citealt{mcnamara07}).   In this  case, one would expect
cooling  to  proceed  in  the  regions unaffected  by  local  heating.
Recently,   deep  X-ray   observations   of  Perseus   and  A2146   by
(respectively)  \citet{fabian11} and  Russell  et al.~(2012)  revealed
bright  X-ray  filaments  which  alone  could  be  associated  with  a
significant fraction ($\lae 80$\%) of the total inferred cooling rate. In Section 
5 of this paper, we will show a similar result for A2597.

On  the other  hand,  \citet{odea10} studied  a  sample of  7 CC  BCGs
selected on  the basis of an  IR excess associated  with elevated star
formation rates, and found that each tended to possess a weak, compact
radio source. The combination of  higher SFR and lower radio power may
be  consistent with a  scenario wherein  a low  state of  AGN feedback
allows  for increased  residual  condensation from  the ambient  X-ray
atmosphere, accounting  for the  elevated star formation  rates.  This
would be consistent with  scenario (iii), though A2597 does not appear 
to be a strong candidate.   We stress that scenarios (i), (ii), and (iii)
are not contradictory or competitive  with one another, and each could
play a simultaneously  important role. The same is  true for possibility
(iv), which we investigate in more detail below.

\begin{figure}%[!htb]
\begin{center}
\includegraphics[scale=0.37,angle=270]{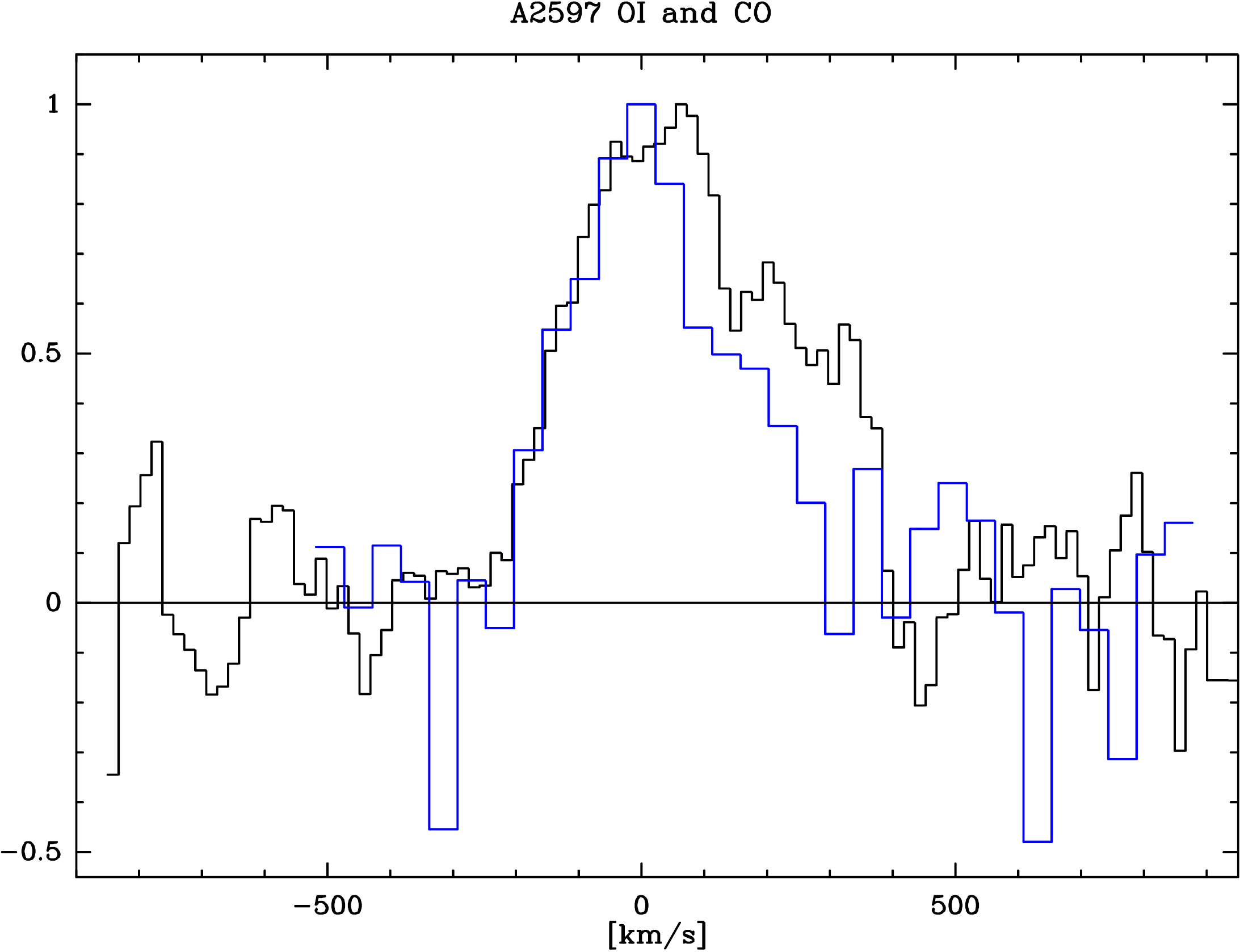}
\end{center}
\caption{A   comparison    of   the   {\it    Herschel}   [\ion{O}{i}]
  $\lambda$63$\mu$m line profile, in black, with the IRAM 30 m CO(2-1)
  rotational line profile in  blue.  The [\ion{O}{i}] line is centered
  at  the  systemic  BCG  redshift,  and  has  an  intrinsic  FWHM  of
  $405\pm55$  km sec\mone\  with  a possible  asymmetric red  velocity
  excess offset from the systemic velocity by $\sim +250$ km sec\mone.
  The CO(2-1) FWHM is  slightly narrower ($300\pm50$ km sec\mone), and
  may also exhibit a slight  red asymmetry. The spectral resolution of
  the CO  data is 45 km  s\mone. Both spectra have  been normalized to
  their maximum to enable comparison of the line profiles. Intensities
  are therefore arbitrary.}
\label{fig:kinematics}
\end{figure}

\subsection{New results on cold gas kinematics --- discriminating between 
hot and cold accretion scenarios}

\begin{figure*}
\begin{center}
\includegraphics[scale=0.45]{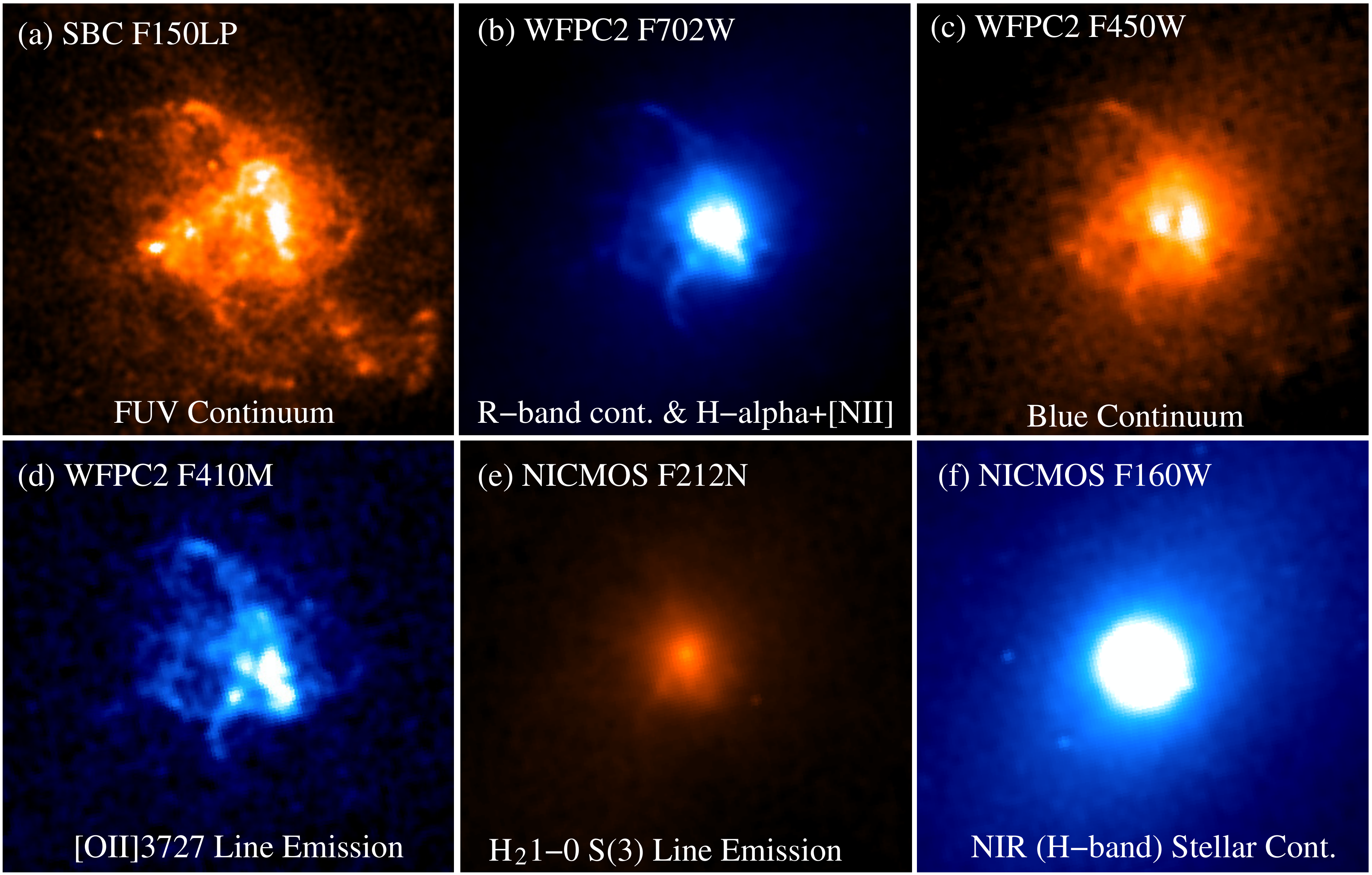}
\end{center}
\caption{{\it  HST} imaging of  FUV, optical,  NIR, and  line emission
  associated with the  $\sim10$ kpc-scale nebula at the  center of the
  A2597 BCG. The FOV of each figure is approximately $10\arcsec \times
  10\arcsec$ ($\sim  15$ kpc$\times 15$  kpc). Panel ({\it a})  is FUV
  continuum emission,  attributed to ongoing star  formation, from the
  ACS SBC F150LP observation of \citet{oonk11}. Ly$\alpha$ emission is
  not included  in the bandpass.  Panel ({\it b}) is  $R$-band optical
  continuum,  H$\alpha$+[\ion{N}{ii}], and  [\ion{S}{ii}]  emission from
  \citet{holtzman96}. Panel  ({\it c}), also  from \citet{holtzman96},
  contains  blue  optical  continuum  and a  small  contribution  from
  [\ion{O}{ii}]$\lambda  3727$  \AA\   emission,  which  dominates  the
  bandpass  in  the  F410M  image  shown  in  panel  ({\it  d}),  from
  \citet{koekemoer99}. Panel ({\it e})  is primarily emission from the
  1.956 $\mu$m (rest frame)  1-0 S(3) H$_2$ line, originally published
  by \citet{donahue00}.  Panel ({\it f})  is {\it H}-band  NIR stellar
  continuum emission, also from \citet{donahue00}.  }
\label{fig:a2597nebulamosaic}
\end{figure*}

The new  {\it Chandra} and {\it  Herschel} data can be  used to better
understand  the origin  of the  $1.8 \pm  0.3 \times10^9$  \Msol\ cold
reservoir in  A2597. Relative  to BCGs in  non-cool core  clusters, CC
BCGs are far more likely to harbor central emission line nebulae, cold
gas components, and  radio sources (e.g., \citealt{mittal09,hudson10},
and references therein).  This strong circumstantial evidence suggests
a  causal connection  between these  warm/cold phenomena  and residual
condensation from the  ICM, but the literature has  so far not reached
consensus as  to the relative  importance of gas-rich major  and minor
mergers  in  this  relationship (e.g., \citealt{sparks12}).   \citet{koekemoer99},  for  example,
suggest that some physical features of the A2597 emission line nebula
are  consistent with  what might  be  expected from  a gas-rich  minor
merger. The  BCG stellar  isophotes do not  show any obvious  signs of
disturbance, so  a recent major merger  is likely ruled  out, at least
within the past several dynamical times.

In  Fig.~\ref{fig:kinematics}  we  plot the  [\ion{O}{i}]  $\lambda$63
$\mu$m  line  profile  from   the  {\it  Herschel}  PACS  spectroscopy
\citep{edge10spec} in black, over  the CO(2-1) rotational line profile
(in blue) from the IRAM  30 m observations of A.~Edge and P.~Salom\'{e}
(pvt.~comm.~2012,   data    previously   unpublished).    [\ion{O}{i}]
$\lambda$63$\mu$m  is one  of  the primary  atomic  coolant lines  for
$T\lae40$  K gas \citep{kaufman99},  and CO(2-1)  can be  considered a
tracer    of   the    molecular    hydrogen   reservoir.     Following
\citet{solomon97} and  adopting a standard Milky  Way H$_2$ mass-to-CO
luminosity ratio of $\alpha=4.6$  \Msol\ (K km s\mone\ pc$^2$)$^{-1}$,
the CO line  luminosity of $L^\prime_\mathrm{CO} = 3.9  \times 10^8$ K
km s\mone\ pc$^2$ converts to a CO-inferred molecular hydrogen mass of
$M_{\mathrm{H}_2}=1.8\pm0.3 \times10^9$ \Msol.

[\ion{O}{i}]  was detected  at  S/N$\gae$30 but  unresolved (the  {\it
  Herschel} beam  at this wavelength is $\sim 9\arcsec  \approx 13.5$
kpc).  The [\ion{O}{i}] line is centered at
the systemic BCG redshift, and  has an intrinsic FWHM of $405\pm55$ km
sec\mone\ with  a possible asymmetric red velocity  excess offset from
the systemic velocity by $\sim +250$ km sec\mone.  The CO(2-1) FWHM is
slightly narrower ($300\pm50$ km sec\mone) though the profile may also
exhibit   a  scaled  down   signature  of   the  same   red  asymmetry
\citep{edge10spec}. The IRAM 30 m CO(2-1) beam is $13\arcsec \approx 20$ kpc, larger than the {\it Herschel}  63$\mu$m beam, so a red wing arising from gas near the center would be more diluted in the CO(2-1) profile.  
A  very marginal (2.6$\sigma$)  CO(1-0) detection
was also obtained by A.~Edge and P.~Salom\'{e} at the same velocity as
the CO(2-1)  line, suggesting that  the CO(1-0) data is  beam diluted.
This means  that the CO  is probably concentrated  in a region smaller than 11\arcsec (17 kpc).  The CO line ratio (if correct) is
close to or larger than  4, suggesting this as well.  \citet{odea94HI}
detected an  unresolved narrow \ion{H}{i} absorption  component in the
nucleus  with a  FWHM of  $\sim 220\pm10$  km sec\mone,  as well  as a
broader component spatially resolved on the scale of the 8.4 GHz radio lobes with a FWHM of  $\sim412\pm40$ km sec\mone, consistent with the
{\it Herschel} results.

\citet{donahue00}  detected H$_2$ 1-0 S(3) 2.2
$\mu$m emission stemming from vibrationally excited molecular hydrogen
on  roughly the  same  spatial scale,  with  some resolved  morphology
tracing the  optical emission line  filaments. The emission  is far
too bright  to be  directly accounted  for by a  cooling flow,  and UV
irradiation by the young stellar  component was the only model capable
of accounting for both the H$\alpha$/H$_2$ line ratios, as well as the
FUV  continuum strength. That study ruled out AGN  photoionization, fast  and slow  shocks,  and X-ray
heating  as the dominant ionization source. 
Excitation by non-thermal electrons may play an important role \citep{ferland09}. 
 \citet{jaffe05}   and
\citet{oonk10} have detected molecular gas  out to a maximal radius of
$\sim 20$ kpc northward of the nucleus.  Thus far, this is the largest
known radial  extent at which a  tracer of the warm/cold  ISM has been
detected  in A2597  (the longest  CC BCG  optical filaments  have been
detected  out   to  $\sim  50$   kpc,  for  Perseus/NGC   1275,  e.g.,
\citealt{conselice01,fabian03,hatch06}).   This  does  not imply  that
there  is no  warm/cold  gas beyond  this  boundary, as  all of  these
observations  are limited by  sensitivity.   The  average
velocity  profiles  for the  ionized  and  molecular  gas, studied  by
\citet{oonk10} using  ESO Very Large Telescope (VLT) SINFONI data (see Table 1), are comparable  to one another
and are centered at roughly  the systemic velocity. The data show smaller scale 
velocity (and velocity  dispersion) excesses that
appear to be  associated with the propagation of  the radio source \citep{tremblay_feedback}.

\begin{figure*}
\begin{center}
\includegraphics[scale=0.75]{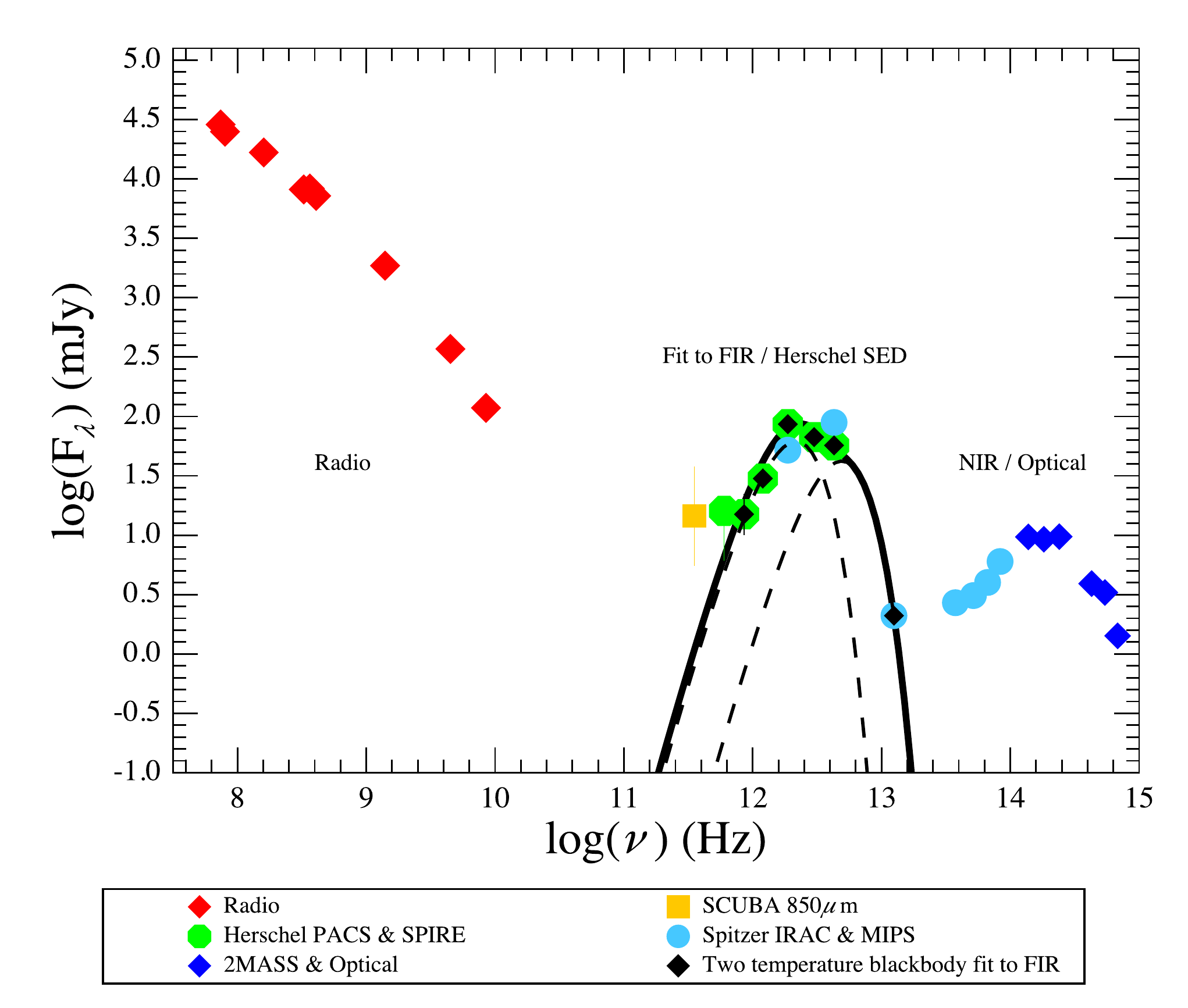}
\end{center}
\caption{Spectral energy distribution of A2597, from radio (left side)
  through optical  (right side)  frequencies. References for  the data
  points     used    in     this    plot     can    be     found    in
  Table~\ref{tab:observations}.  Following the  method from  \citet{mittal11},  a two-temperature blackbody has been  fit to the
  FIR portion  of the  SED, sampled by the new  {\it Herschel} PACS  and SPIRE
  observations. See Section \ref{section:herscheldust} for a discussion of this figure. The 500 $\mu$m SPIRE point is an upper limit. Because the plot is log-log, the error bars on most points are smaller than the points themselves. }
\label{fig:sed}
\end{figure*}

Gas cooling from an ambient X-ray atmosphere should be associated with
low  net  angular  momentum.  Its velocity  profile  should  therefore
smoothly increase  from the systemic  velocity at larger  radii (e.g.,
$\sim 50$  kpc) to  a few hundred  km sec\mone\  in the center  of the
BCG. As discussed above, this is  consistent with what has so far been
observed.  On the  other hand, cold gas acquired  through a merger may
be  associated  with higher  net  angular  momentum  and exhibit  bulk
rotation   at  larger   radii.   The   relative  consistency   of  the
[\ion{O}{i}], CO(2-1),  \ion{H}{i}, and H$_2$  velocity profiles seems
consistent  with  what  would  be  expected if  residual  ICM  cooling
deposited  these  gaseous components  into  the  BCG  with low  excess
angular momentum or velocity structure.   While we cannot speak to the
velocity structure of the X-ray gas on these scales, the fact that the
X-ray isophotes are consistently elongated along the major axis of the
BCG  (top right panel of Fig.~\ref{fig:chandra})  supports the notion that the
two components are roughly similar in terms of angular momentum.

Alternatively,  a   merger  origin  for   the  gas  cannot   be  ruled
out. Although  high velocity gas is  not detected at  large radii, and
although the cold gas is concentrated in the very inner regions of the
nucleus without complex or high amplitude velocity structures, one can
imagine a merger  producing similar results if enough  time has passed
since the merger for the gas  to reach the nucleus and couple with the
local dynamics. A possible merger origin for the warm/cold gas phases
in A2597 is discussed in detail by \citet{koekemoer99}.

A  purely merger-based  origin  for  this gas  would  be difficult  to
reconcile with the {\it  FUSE} and {\it XMM-Newton} results consistent
with cooling  gas $<10^6$  K, 
unless conductive interfaces produce the lines attributed to cooling (e.g., \citealt{sparks12}). 
The  observed star
formation X-ray  cooling time threshold \citep{rafferty08,cavagnolo08}, 
which we will discuss in Section 5, would also be difficult to understand 
in a ``mergers only'' scenario. 
In the  sections below  we will
argue that  ICM contributions likely  play a very significant  role in
A2597, regardless of the  unknown contribution by cold gas acquisition
scenarios  like mergers.  We reiterate  that a  recent  gas-rich major
merger is almost certainly ruled out for A2597.

\subsection{New Herschel constraints on the cold gas and dust components}

\label{section:herscheldust}

The  $1.8 \pm 0.3 \times10^9$   \Msol\  cold  molecular  gas
reservoir in A2597  is cospatial with the $\sim  10$ kpc emission line
nebula, which  has been studied  extensively in the  literature (e.g.,
\citealt{heckman89,mcnamara93,voit97,mcnamara99,koekemoer99,oegerle01,odea04,jaffe05,oonk11}).
High  spatial   resolution  archival  {\it  HST}   imaging,  shown  in
Fig.~\ref{fig:a2597nebulamosaic},  reveals   its  complex  filamentary
morphology in the  FUV and optical (panels {\it a}  through {\it d} in
Fig.~\ref{fig:a2597nebulamosaic}).    Deep  optical   spectroscopy  by
\citet{voit97}  shows  the gas  temperature,  abundance, and  electron
density of  the nebula  to be 9,000-12,000  K, $\sim0.5  Z_\odot$, and
$n_e \sim 200$ cm$^{-3}$, respectively.  \citet{koekemoer99} estimated
the pressure associated with the $10^4$ K gas on these scales to be
\begin{equation} 
p_{\mathrm{nebula}} \approx 5.5 \times 10^{-10} \left(\frac{n_e}{200~\mathrm{cm}^{-3}}\right)\left(\frac{T}{10^4~\mathrm{K}}\right)~\mathrm{dyn~cm}^{-2}.
\end{equation}
Our estimated central  X-ray pressure on the same  $\sim 10$ kpc scale
is $7\times10^{-10}$ dyn cm\mtwo\ \citep{tremblay_feedback}, very close to what is inferred for the 10$^4$ K
gas, implying that the two phases are in rough pressure equilibrium.

The nebula is associated with a substantial dust component that can be
quantitatively  studied  using the  {\it  Herschel} observations.   In
Fig.~\ref{fig:sed}  we show the  radio through  optical SED  of A2597,
with the  FIR component filled in  by the new {\it  Herschel} PACS and
SPIRE data (green circles). Following the same method described in \citet{mittal11}, a
two-temperature  black body  is fit  to the  {\it Herschel}  (PACS and
SPIRE)  and  {\it  Spitzer}  24   $\mu$m  data  points  to  model  the
contribution by dust to the FIR SED. 
Our fit assumes a dust absorption coefficient of $\kappa_\nu = 5.6 \times \left(\nu / 3000~\mathrm{GHz}\right)^\beta$ m$^{2}$ kg\mone\
with a dust emissivity index of $\beta=2$ \citep{dunne00,mittal11}. We note that the 850 $\mu$m SCUBA
data  point could  also be  associated  with an  additional cold  dust
component, though  we choose not to  include it in  our modeling given
the uncertain radio source contribution to this flux.

We find  the best-fit  temperature  and mass  of the warm  dust component  to be
$T_\mathrm{warm~dust}       =      47\pm      1.4$       K      and
$M_\mathrm{warm~dust} = \left(1.7\pm0.6\right)   \times  10^5$  \Msol,
respectively.  The temperature and mass  of the cold dust component is
found     to    be     $T_\mathrm{cold~dust}=20\pm1.7$     K    and
$M_\mathrm{cold~dust} = \left(1.3\pm0.5\right)\times10^7$        \Msol,
respectively.   
This two component  blackbody fit could be  unrealistic if
(1) there is  a large population of small dust  grains driving a broad
emission distribution  that might  mimic a modified  blackbody, and/or
(2)   if   the   warm    dust   grains   are   heated   stochastically 
(which is only realistic for very small grains, e.g., \citealt{draine01}). 
Furthermore, a two-temperature fit assumes that there is no 
dust at temperatures between $\sim 20$  K and $\sim 48$ K, which is 
almost certainly not true.  
The SPIRE and SCUBA points suggest the presence of very cold dust, which 
would increase the mass limit. Finally,   
the best-fit dust
mass is extremely sensitive to the assumed dust temperature, which in turn is degenerate with 
the dust emissivity index $\beta$. We have assumed that $\beta=2$, but the value can 
vary between $\sim 1.5$ and 2 (e.g., \citealt{tabatabaei11}). 
While we find that varying $\beta$ between these values does not significantly 
change our results, we stress that the dust masses and temperatures 
listed here are extremely assumption-heavy. Altering these assumptions even slightly can significantly 
alter the result. As a point of illustration, the single-temperature
fit to the A2597 {\it Herschel} points by \citet{rawle12} finds $T=31$ K and $M_\mathrm{dust}=1.6\times10^8$ \Msol, 
an order of magnitude higher than our result.  
Ultimately, we lack the data to speak affirmatively about the validity of these many assumptions.

With these  disclaimers noted,  our total {\it  Herschel}-derived dust
mass of  $M_\mathrm{cold~dust} = 1.3\times10^7$  \Msol\ compared
to    the    inferred    molecular    gas   mass   of
$1.8 \pm 0.3 \times10^{9}$  \Msol\ implies a gas-to-dust ratio of $\gae  140$. 
We again  stress that,  while {\it  Herschel} can  place the
tightest available constraints on the A2597 dust mass, this gas-to-dust
ratio remains an uncertain lower-limit given  
(1) the dust mass, temperature, and fit parameter degeneracies discussed above, and (2)  
CO-inferred molecular gas masses are
inherently uncertain  because of the CO-H$_2$  conversion ``X factor''
(e.g., \citealt{liszt10}, and  references therein).  
This is especially true considering the apparent low metallicity of the gas, 
which could mean that our assumed CO-H$_2$ conversion factor is an order of magnitude too low (e.g., \citealt{bolatto11}). We could therefore be significantly underestimating the gas mass.

Nevertheless, the
inferred lower-limit ratio is comparable to Galactic ratios, and somewhat low relative to past
estimates for other  CC BCGs (e.g., \citealt{edge01}).  These older  ratios suffer  from an
inability  (at the time)  to place  convincing limits  on CC  BCG dust
masses,  so a  more comprehensive  study with  the new  {\it Herschel}
sample  is needed  to  better understand  the ``typical''  gas-to-dust
ratios of  CC BCGs  (assuming such a  typical ratio exists  --- it
certainly  may  not).  Unpublished  and  science  demonstration  phase
results  for the {\it  Herschel} CC  BCG sample  so far  indicate more
MW-type   gas-to-dust  ratios   on   average  (e.g., \citealt{edge10spec,edge10phot,rawle12}; Oonk et al.~2012 in prep).

The   {\it  Herschel}  [\ion{C}{ii}]-to-FIR  luminosity
ratio, which  is a tracer  of the relative  cold gas and  dust cooling
rates,  is $\sim  0.014$ for  A2597. This  value is  high  relative to
similarly    FIR    luminous     star    forming    galaxies    (whose
[\ion{C}{ii}]/L$_{\mathrm{FIR}}$  ratios  are  generally  an  order  of
magnitude lower).  The ratio seems  to be higher for lower metallicity
galaxies (e.g., \citealt{maiolino09},  and references therein), so the
relatively high  ratio estimated for  A2597 could be partially  due to
lower metallicity gas, consistent with a cooling flow scenario.

The MW-type gas-to-dust ratio means that the star forming 
gas in A2597 is quite dusty. This is inconsistent with a scenario 
wherein already dusty gas originates in the hot atmosphere, where  grain sputtering  timescales are  short (e.g.,
\citealt{draine79}, review by  \citealt{draine03}).  In this case, one
might   expect   far  higher   gas-to-dust   ratios (i.e., dust-poor gas) stemming from cooling  flow origin  scenarios.  Furthermore, even if  the
cooling  flow gas were  dusty despite  short grain  destruction times,
then extended FIR emission might be expected on the same scales as the
X-ray gas, as  dust emission should be a major  channel of energy loss
for the cooling ICM.  This extended FIR emission cannot be detected in A2597 by  {\it Herschel} due to spatial resolution limitations
\citep{edge10phot,edge10spec}.

\begin{figure*}
\begin{center}
\includegraphics[scale=0.55]{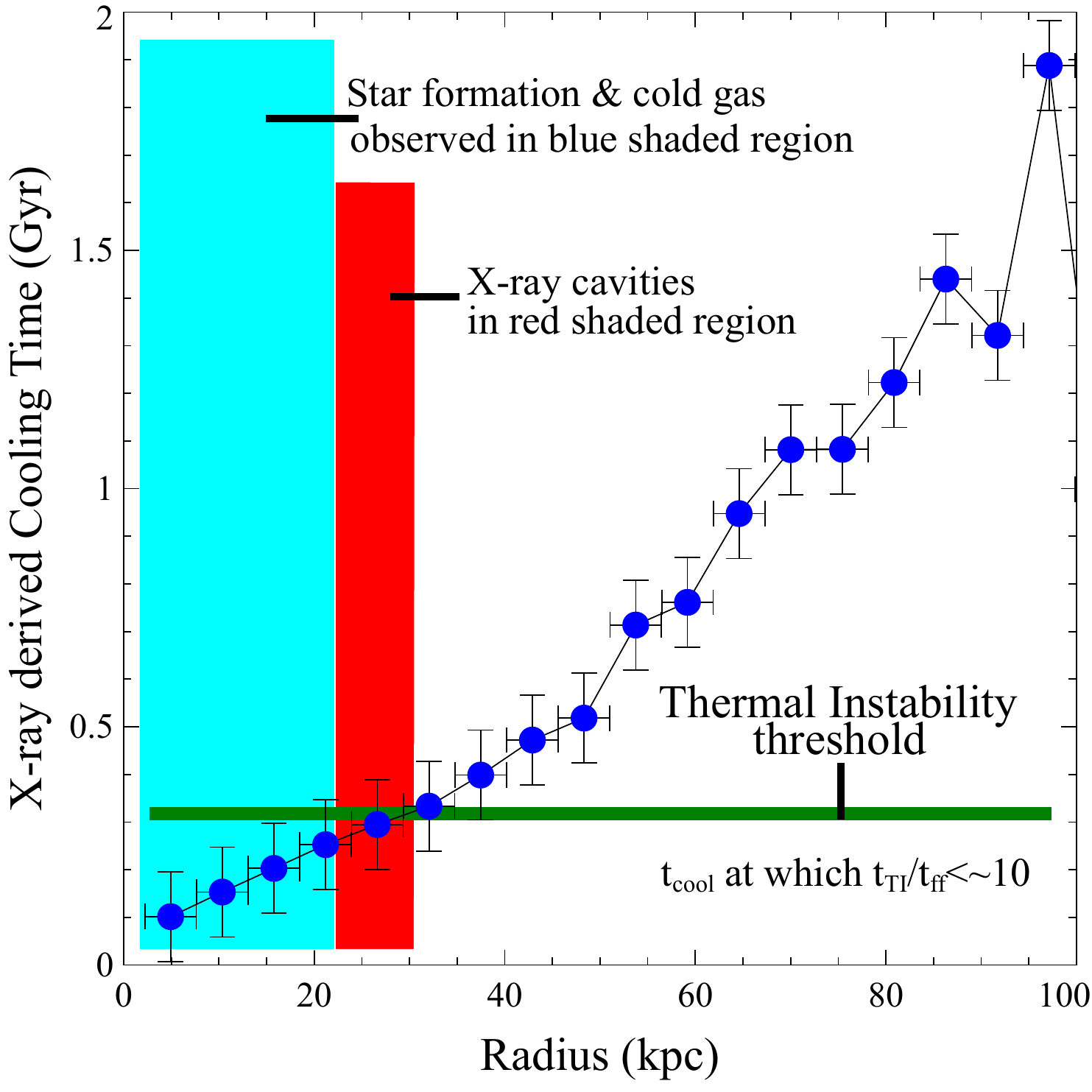}\includegraphics[scale=0.55]{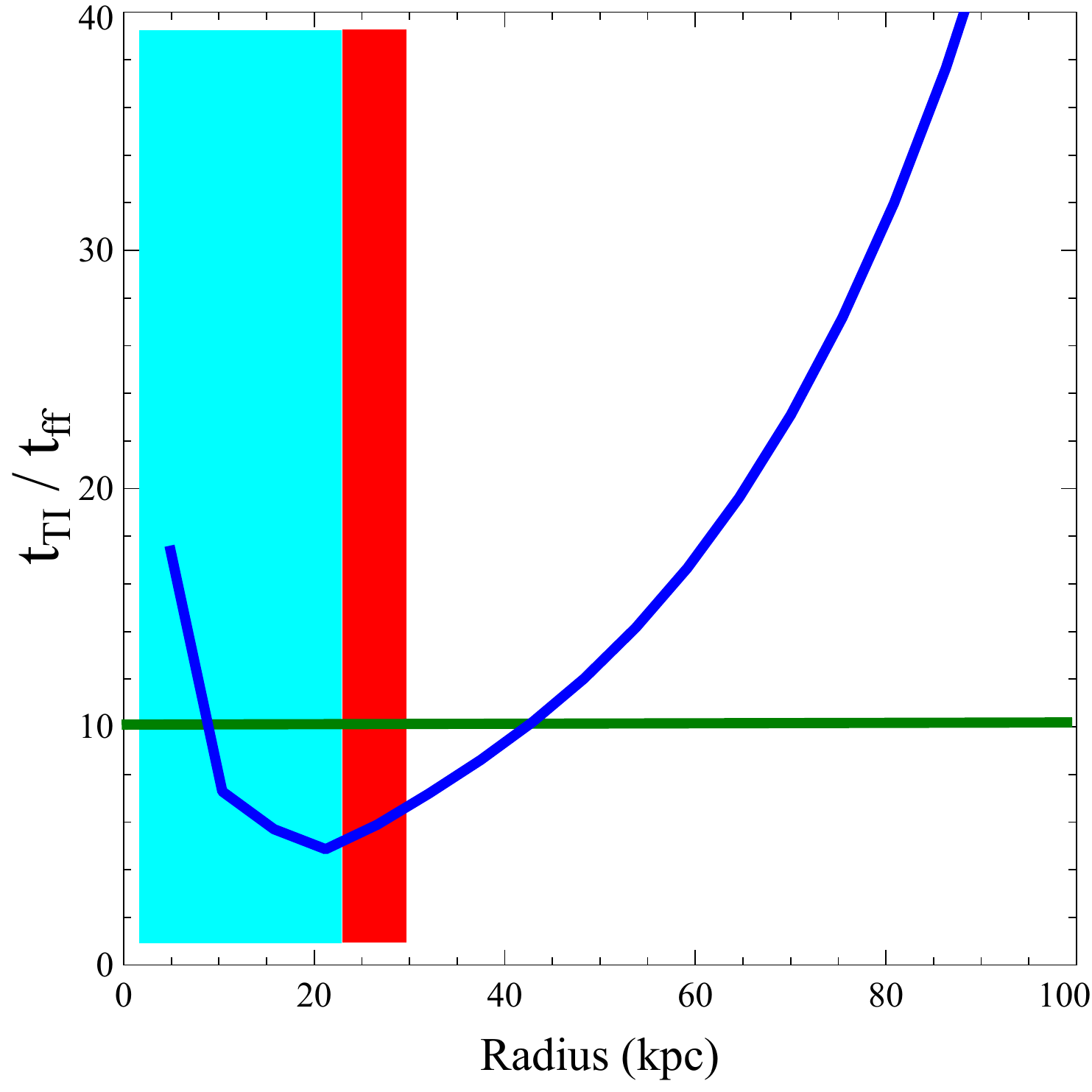}
\end{center}
\caption{({\it left})  X-ray cooling time  profile vs.~cluster-centric
  radius.         The        green        line        marks        the
  $t_{\mathrm{TI}}/t_{\mathrm{ff}}\lae    10$    thermal   instability
  threshold  discussed by  \citet{sharma11}, where  $t_\mathrm{TI}$ is
  the thermal  instability timescale and $t_\mathrm{ff}$  is the local
  gravitational  freefall timescale.  \citet{sharma11}  suggests that,
  beyond this  threshold, a cooling flow can  form thermally unstable,
  star forming clouds and filaments.  The blue shaded region marks the
  largest radial extent of observed ionized and molecular gas. The red
  shaded region, which overlaps it, marks the largest radial extent at
  which  X-ray cavities  are  observed.  ({\it  right})  Ratio of  the
  thermal    instability    timescale    and    freefall    timescales
  vs.~cluster-centric            radius.            The           same
  $t_{\mathrm{TI}}/t_{\mathrm{ff}}\lae 10$ star formation threshold is
  marked with  a green line.   See Section \ref{section:sharma}  for a
  discussion of these profiles. }
\label{fig:sharma}
\end{figure*}

The above result suggests that while the cooling ICM may 
contribute the bulk of the cold gas in A2597, it cannot 
contribute the substantial dust component associated with that 
cold gas. Instead it is more likely that the bulk of the dust 
component is contributed by stellar mass loss (e.g., via dust-rich 
AGB winds and supernovae). Modern cooling flow models 
suggest that cooling from the ICM concentrates in thermally 
unstable clouds and filaments (e.g., \citealt{fabian11}).  In   these
thermal instabilities,  the cooling/cold gas may shield  the dust from
sputtering. The recent detection  by \citet{donahue11} of PAH emission
in  A2597 almost  certainly necessitates  such a  shielding mechanism,
given  the   extreme  fragility  of  PAH  molecules   in  hot  ambient
environments. 
If the dust is not efficiently shielded within the filaments, then
dust production rates must
be higher than what is implied by most 
models for dust production.

\citet{voit11} argue that mass loss from the old stellar population
of the BCG is indeed a significant source of dusty 
gas in many cool core clusters. We note that this scenario is not at odds
with the residual cooling flow model for A2597.
As filaments condense
from the cooling hot atmosphere, they do so within the stellar body 
of the galaxy and should therefore contain existing stars within 
their volume.
\citet{voit11} suggest 
that the ejected dust-rich envelopes stemming from the old stellar component are not assimilated into the hot phase, but rather remain cold and confined to the cold ambient clouds and filaments, where the dust may be shielded from interaction with the hot phase.  
If the old stellar component plays an important role in enriching the filaments with 
dust, then it is likely that it at least plays a non-negligible role in contributing 
to the observed cold gas mass as well.

\begin{figure*}
\begin{center}
\includegraphics[scale=0.5]{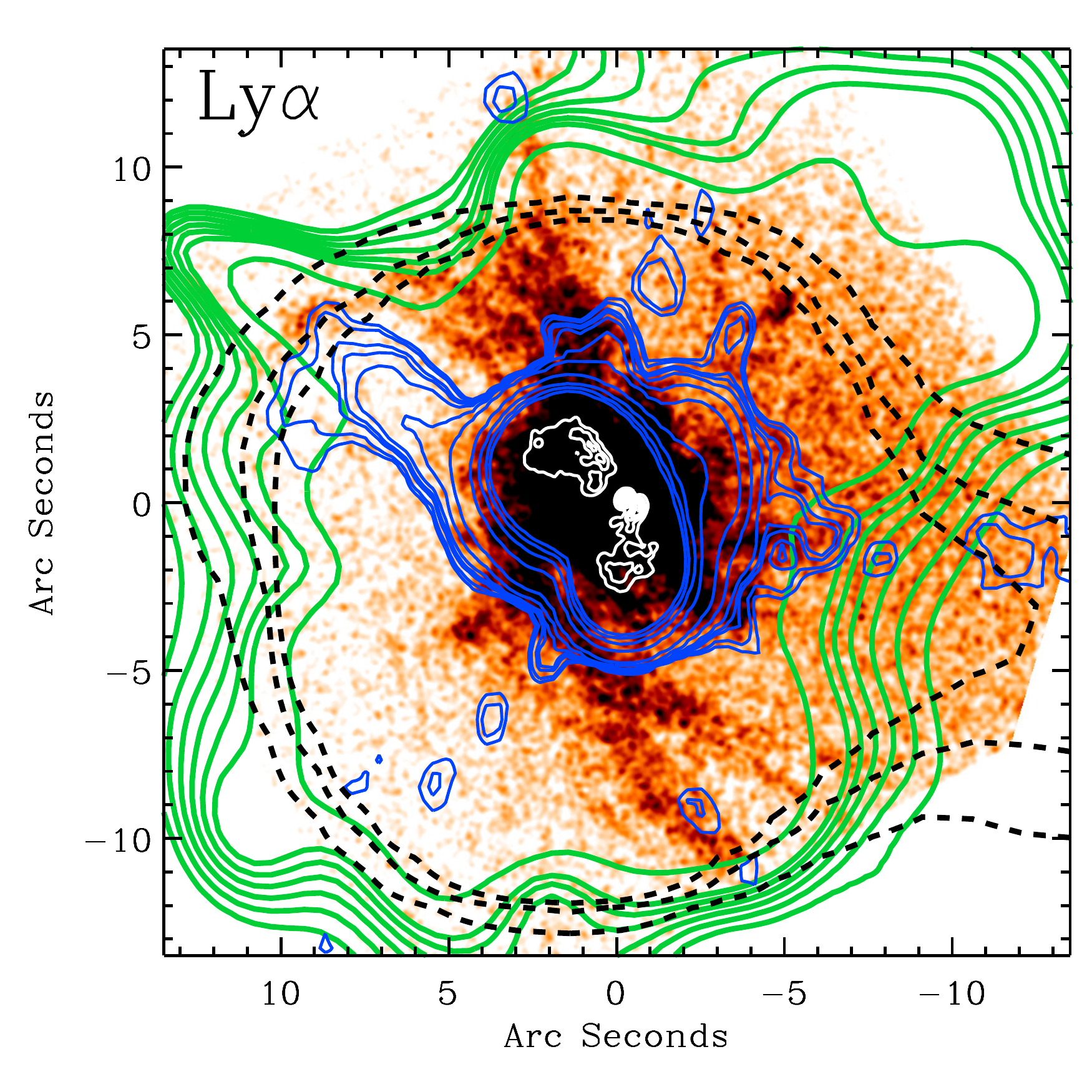}\includegraphics[scale=0.5]{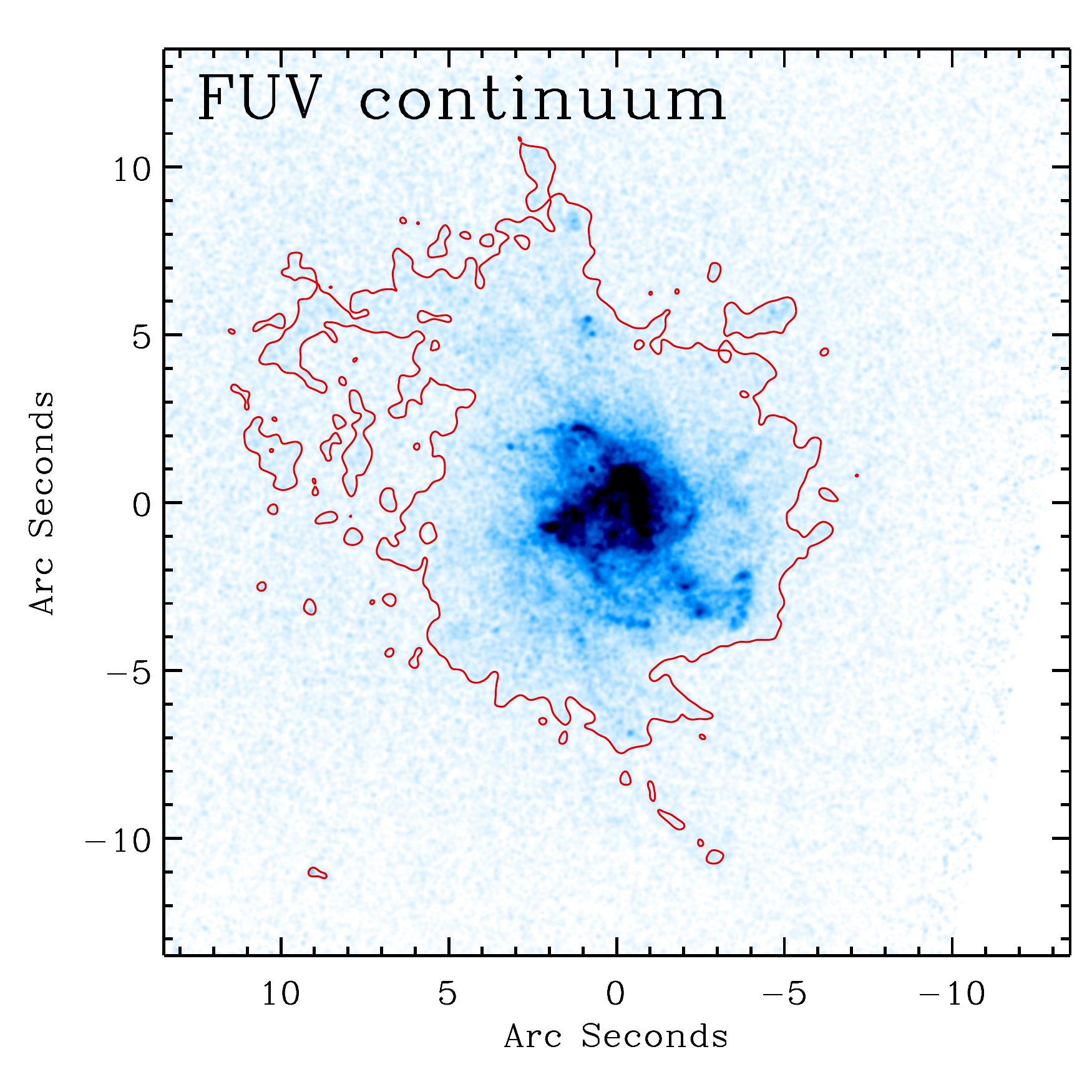}\\ \hspace*{7mm}
\includegraphics[scale=0.65]{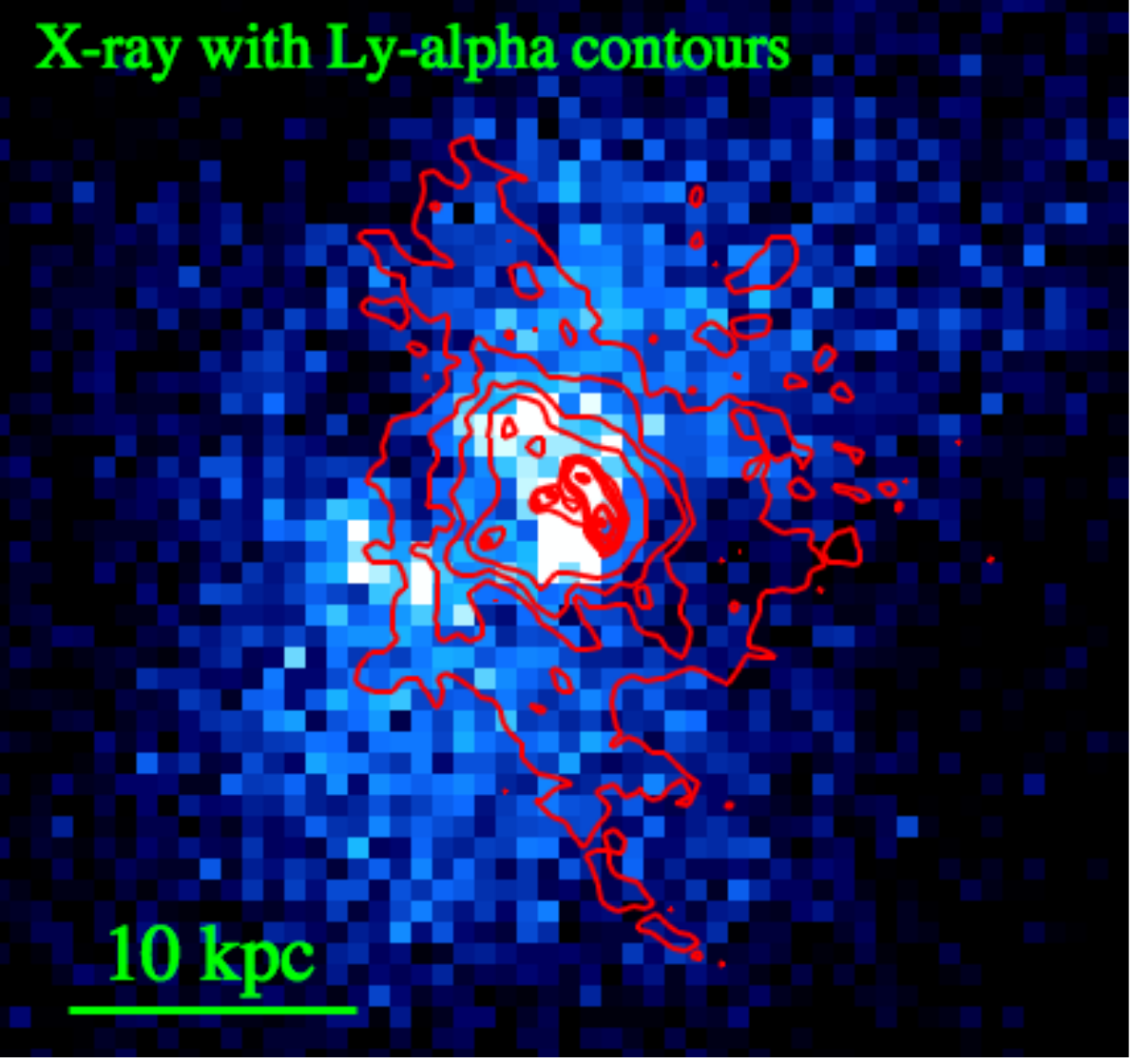}\hspace*{5mm}
\includegraphics[scale=0.65]{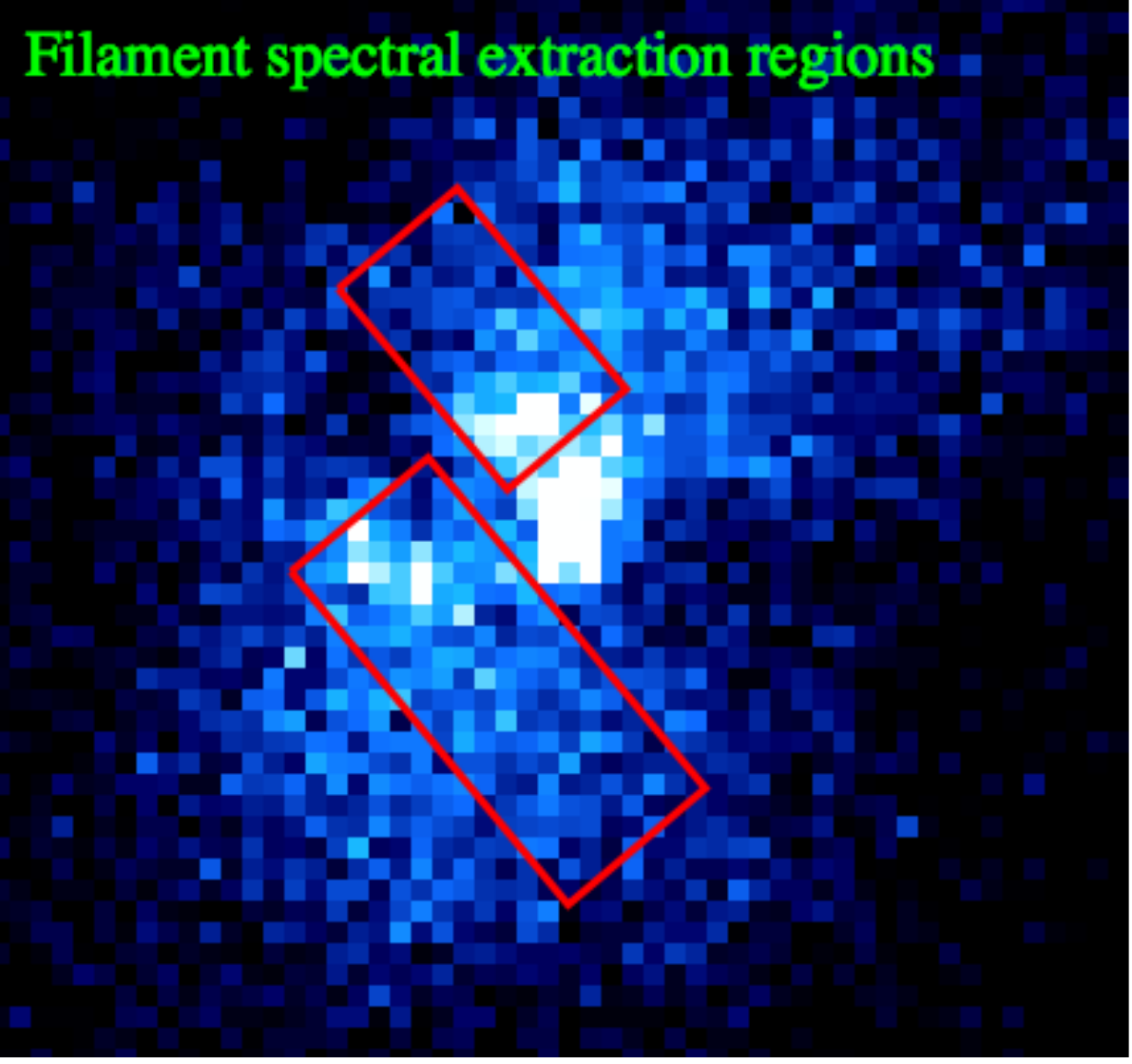}
\end{center}
\caption{({\it  top left}) In red/orange  we show  the {\it  HST}/STIS FUV
  observation  of  extended Ly$\alpha$  emission  associated with  the
  A2597 BCG, from \citet{odea04}. 1.3  GHz, 8.4 GHz, and 330 MHz radio
  contours  are   overlaid  in   blue,  white,  and   black  (dashed),
  respectively, while adaptively smoothed 0.5-7 keV X-ray contours are
  overlaid in green.  We have  removed the innermost 1.3 GHz, 330 MHz,
  and  X-ray  contours  to  aid viewing.   ({\it top right}) {\it  HST}/ACS SBC F150LP
  $\sim 8$ ksec exposure of FUV continuum emission associated with the
  A2597 emission line nebula.  In red contours, we outline the
  boundary within which low  surface brightness FUV continuum emission
  has been detected  at $\sim 2\sigma$ above the  background.   ({\it bottom left}) The unsmoothed 150 ksec 0.5-7 keV X-ray data, shown with 
an aggressive color scale stretch used to highlight the high surface brightness knots observed near the X-ray centroid. Ly$\alpha$ contours 
are overlaid in red. The northern high surface brightness X-ray knot bends upwards to follow the overall distribution of the high surface brightness Ly$\alpha$ and FUV continuum emission. 
({\it bottom right}) The same X-ray image, with red boxes marking the sectors from 
which X-ray spectral data were extracted and modeled. A \texttt{WABS}$\times$\texttt{MKCFLOW} fit to the data finds a combined mass deposition rate of $\sim 7 \pm 0.5$ \Msol\ yr\mone\ associated with the X-ray filaments. This is comparable to the local star formation rate estimated
from FUV emission in the filaments. }
\label{fig:fuvlya}
\end{figure*}

\section{New Results on Star Formation}

\subsection{{\it Herschel} constraints on the star formation rate}

The $8-1000~\mu$m luminosity can be used with the Kennicutt relation to estimate the FIR-inferred star formation rate \citep{kennicutt98}; \begin{equation}
\frac{\mathrm{SFR}}{M_\odot~\mathrm{yr}^{-1}} \lae 4.5 \times \left( \frac{L_{\mathrm{FIR}}}{10^{44}~\mathrm{ergs~s}^{-1}} \right).
\end{equation}
Using $L_{\mathrm{FIR} } = \left(6.50 \pm 1.35 \right)\times10^{43}$ ergs s\mone\ ($1.69\pm0.35 \times 10^{10}$ $L_\odot$) derived from
the  {\it  Herschel}  data,   we  find  SFR$_\mathrm{FIR}  \lae  2.93\pm0.69$
\Msol\ yr\mone.  This is consistent  with the many past SFRs estimated
for A2597,  which range from  $2-12$ \Msol\ yr\mone\ depending  on the
method used  \citep{mcnamara93,odea04,donahue07,oonk11,rawle12}.  Estimates of
star formation rates (particularly in the FUV) are extremely sensitive
to internal extinction by dust. The  
Balmer sequence in A2597 suggests that the 
internal extinction is significant (about one magnitude  in $V$-band, i.e.,  $A_V\sim 1$, \citealt{voit97,oonk11}), consistent with the substantial dust mass 
inferred from the {\it Herschel} data.

\subsection{Star formation entropy threshold}

\label{section:sharma}

Using  numerical  simulations and ignoring thermal conduction,   \citet{sharma11}  showed  that  local
thermal  instabilities   in  a  cooling  flow  will   only  produce  a
multiphase, star forming ISM when the ratio of the thermal instability
timescale  $t_\mathrm{TI}$   to  the  local   gravitational  free-fall
timescale $t_\mathrm{ff}$ is $t_{\mathrm{TI}}/t_{\mathrm{ff}}\lae 10$
(see \citet{sharma11} for definitions of these quantities). 
Expressed in terms of the gas entropy $S$, this thermal instability threshold is $S\sim 20$ keV~cm$^2$. Importantly, this theoretical result is very close to  the observed $S\sim
30$   keV~cm$^2$   star  formation  onset   threshold  discussed   by
\citet{rafferty08,cavagnolo08}.

The $t_{\mathrm{TI}}/t_{\mathrm{ff}}\lae 10$ threshold is marked in the green line on the X-ray cooling time profile
in Fig.~\ref{fig:sharma}{\it  a} as well  in Fig.~\ref{fig:sharma}{\it
  b},  in which  we plot  the  $t_{\mathrm{TI}}/t_{\mathrm{ff}}$ ratio
vs. cluster-centric  radius. 
In both panels we mark maximal radial extent of (a) Ly$\alpha$, FUV
continuum associated  with star  formation, as well  as warm  and cold
molecular gas inferred from Very  Large Telescope (VLT), {\it HST}, CO
and {\it Herschel} observations and  (b) the X-ray cavity network with
blue and red shaded  regions, respectively. Note that the X-ray cavity network (red region) overlaps with the blue region, meaning it also extends inward to the cluster centre (as can be seen in the bottom two panels of Fig.~\ref{fig:chandra}). 
We calculate the $t_{\mathrm{TI}}/t_{\mathrm{ff}}$  ratio from the X-ray data by 
 fitting third-order polynomials in log space to the temperature and pressure profiles, 
 then analytically differentiating the logarithmic pressure profile to obtain 
 the free-fall time. 
 We account for the presence of the BCG by setting a minimum 
 value for the gravitational acceleration $g$ to be that of a 
 singular isothermal sphere with a velocity dispersion of 250 km s\mone. 
 This BCG contribution is only important at radii $\lae 10$ kpc.

As discussed in \citet{sharma11}, the thermal instability timescale can be approximated as a multiple of the cooling time that is dependent upon 
how much AGN heating one includes in the model.
While the curve in Fig.~\ref{fig:sharma}{\it b} 
is for pure free-free cooling, including a heating component 
could scale it upwards by a factor of $\sim 2$ (for moderate heating) to $\sim 3$ (for strong heating). 
Regardless of whether or not heating is included, the \citet{sharma11}  predictions  are roughly   consistent  with   observations  in   the  case   of  A2597. 
The same is  true of the 30 keV cm\mtwo\ star formation
entropy  threshold  discussed  by  \citet{rafferty08} and \citet{cavagnolo08} --- we
don't mark  this threshold, but it  would lie in nearly the same locations as the 
green  lines in  Figs.~\ref{fig:sharma}{\it a}  and {\it  b}.  
Therefore in either case, the threshold lies just outside the region
containing star formation, ionized and molecular gas, and X-ray cavities.

This  apparent  threshold
would  be  difficult to  understand  if  the cold gas fueling   star
formation was supplied by a merger. One might imagine such a scenario 
if there were a steep radial gradient in thermal conduction efficiency, such 
that clouds introduced by a merger no longer evaporate at the radius where conduction no longer outpaces 
radiative cooling. This is purely speculative, however, 
and the results presented here strongly suggest that a residual cooling flow from the 
ambient hot atmosphere is the primary contributor of the warm and cold gas in A2597. 
We provide further supporting evidence for this suggestion in the following section.

\begin{figure*}
\begin{center}
\includegraphics[scale=0.42]{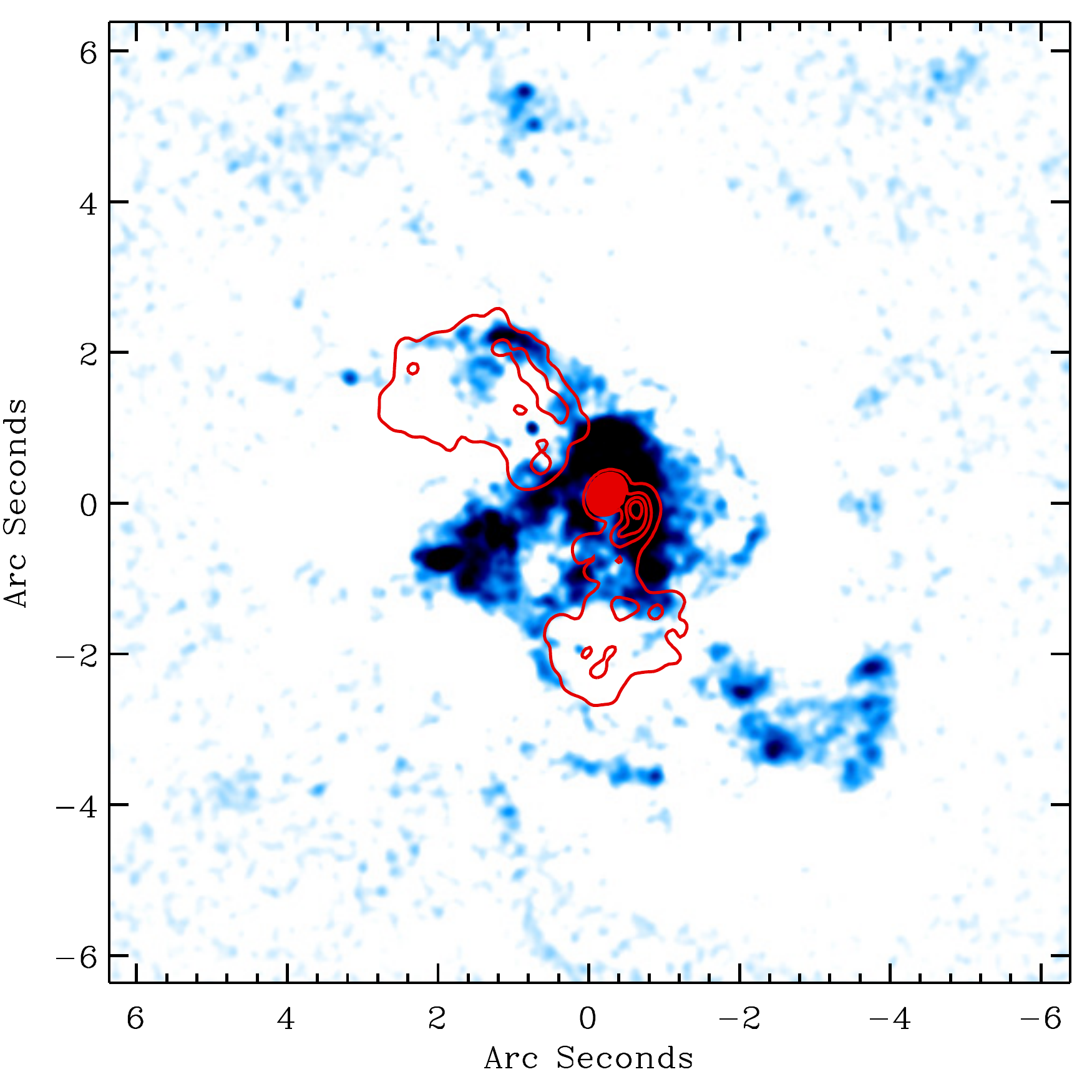}\includegraphics[scale=0.485]{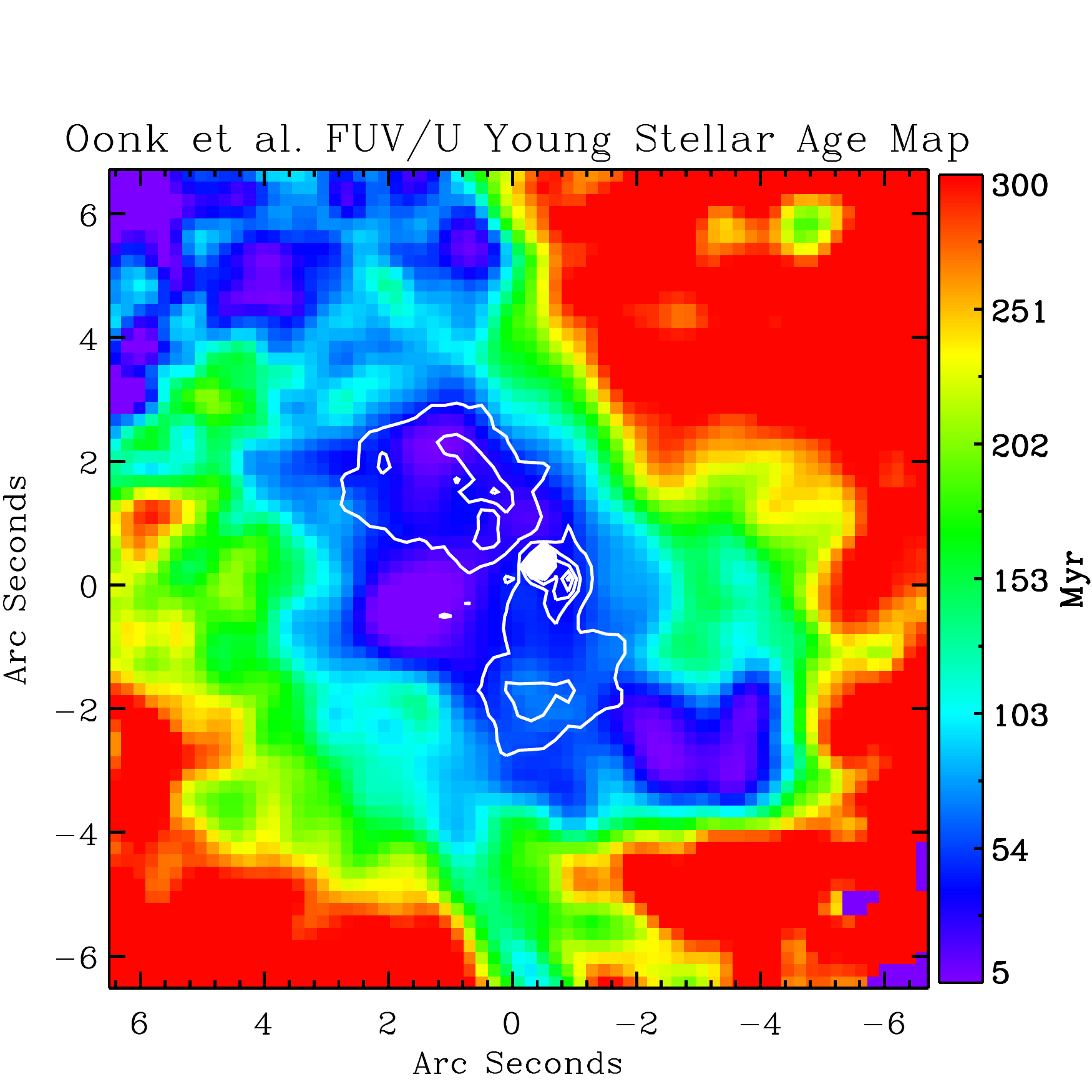}
\end{center}
%\plottwo{final_figures/fuv_resid.eps}{final_figures/agemap.eps}
\caption{({\it left}) Unsharp mask of the 8 ksec {\it HST}/ACS FUV continuum 
image shown in Fig.~\ref{fig:fuvlya}. 
Note the strong spatial correspondence of the Northern arc of FUV 
emission with the leading edge of the 8.4 GHz radio lobe, which is 
outlined in red contours. There is also evidence that the southern 
radio lobe has swept out the star forming gas, or has been allowed 
to expand after emerging from the dense gaseous medium.
({\it right})
Single Stellar Population (SSP) age map from \citet{oonk11}, made by comparing observed FUV/$U$-band ratios (from {\it HST} and the VLT, respectively) to those predicted 
from \citet{bruzual03} models. The youngest portion of the young stellar component is found nearer to the central 8.4 GHz radio source, which we overlay in white contours. The young stars also extend Northwards along the region threaded by Ly$\alpha$ and FUV continuum filaments (as can be seen by comparing this figure with Fig.~\ref{fig:fuvlya}).  }
\label{fig:agemap}
\end{figure*}

\subsection{Evidence for filamentary cooling channels}

\label{section:filaments}

In  Fig.~\ref{fig:fuvlya}{\it  a}  we  show the  extended  filamentary
Ly$\alpha$  emission ({\it  HST}/STIS,  from \citealt{odea04}),  whose
brightness can be accounted for by photoionization from the underlying
FUV  continuum emission  stemming  from the  young stellar  component,
shown   in   Fig.~\ref{fig:fuvlya}{\it   b}   ({\it   HST}/ACS,   from
\citealt{oonk10}).   In the left panel, we show the adaptively smoothed 
0.5-7 keV X-ray contours in green (with the central contours removed 
to aid viewing), and the 330 MHz, 1.3 GHz, and 8.4 GHz radio contours 
in black, blue, and white, respectively. In the right panel we use red contours to mark 
the outermost boundary  at which low surface brightness FUV 
continuum emission is observed at a level of 2$\sigma$ above 
the background.

The faint underlying FUV continuum follows the extended bright 
Ly$\alpha$  clumps  and filaments northwards and southwards, out  
to a cluster-centric radius of nearly 20 kpc. In the X-ray temperature 
map presented in \citet{tremblay_feedback}, there is an elongated region of 
cold X-ray gas extending along the same N-S projected axis. 
This filamentary axis is also significantly offset in position angle 
from the cavity/radio ``AGN heating axis''. 
The Northeastern Ly$\alpha$ filaments lie along a rim
bordering the cold X-ray filament (feature 5 on Fig.~\ref{fig:chandra}), which could be due to the star forming gas having been swept 
outward by the 1.3 GHz radio source (or by the soft X-ray filament that the radio source may have 
dredged upwards, e.g., \citealt{tremblay_feedback}).

In the  bottom two panels  of Fig.~\ref{fig:fuvlya} we show  the 0.5-7
keV  unsmoothed X-ray  data, with  an aggressive  color  scale stretch
applied to  enhance the contrast. Ly$\alpha$ contours  are overlaid on
the bottom left panel.  Note  how the Northern bright X-ray knot bends
5 kpc upwards to follow the high surface brightness Ly$\alpha$ and FUV
continuum  emission.  The  red  boxes  in the  bottom  right panel  of
Fig.~\ref{fig:fuvlya}  right  panel mark  the  sectors  from which  we
extract and  model X-ray  spectral data.  These  sectors approximately
cover the region over which  the extended Ly$\alpha$ and FUV continuum
filaments are cospatial with the soft X-ray gas. As stated previously,
this is also the axis along which elongated cooler X-ray gas is observed
in the X-ray temperature map. 
Following the method described by \citet{tremblay_feedback}, a 
cooling flow model (\texttt{WABS}$\times$\texttt{MKCFLOW}) fit  to the spectra  extracted from
these sectors  finds a  combined mass deposition  rate of $\sim  7 \pm
0.5$ \Msol\ yr\mone\ associated with the region of X-ray gas cospatial
with the  ionized filaments filaments.  This is roughly  comparable to
the  local star  formation rate  estimated  from FUV  emission in  the
filaments.   A   multi-component  fit  to  the   X-ray  spectral  data
\citep[e.g.,][]{sanders04}  suggests  that  multiphase  gas  could  be
present, though the quality of  the fits prevents any conclusions from
being  made.   That these  regions  could  account  for a  significant
fraction of the residual  cooling luminosity is consistent with recent
similar results on Perseus and Abell 2146 from \citet{fabian11} and Russell et al.~(2012), respectively.

This  result serves  as  further supporting  evidence  to our  overall
suggestion that (a) ICM  cooling contributes a significant fraction of
the cold  gas budget fueling the  star formation and  (b) this cooling
happens  in a  spatially discrete,  structured manner  concentrated in
filaments.  In  the case of A2597, these  filaments may preferentially
be  found in a  cooling channel  that is  nearly perpendicular  to the
projected  AGN heating  axis  permeated by  X-ray  cavities and  radio
emission.  Future Atacama  Large Millimeter/submillimeter Array (ALMA)
observations capable of spatially  resolving molecular gas emission on
the  scale of  the filaments  will be  necessary to  either  refute or
reinforce the above claims.

\subsection{Evidence for both triggered and persistent star formation amid AGN feedback}

The frequently studied morphological correspondence  of blue  excess filaments
with  the 8.4 GHz  radio source in A2597  presents an important test  case for
models of  jet-induced star formation. The  morphological evidence can
be seen  in the leftmost  panel of Fig.~\ref{fig:agemap}, in  which we
show an unsharp  mask of the FUV continuum  emission. The leading edge
of the northern  8.4 GHz radio lobe, shown  in red contours, spatially
correlates with the  northernmost FUV arc, which might  be expected if
star  formation is triggered  by shock-induced  cloud collapse  as the
propagating plasma  entrains and displaces cold gas  phases (see e.g.,
the      shock/jet-induced      star      formation     models      by
\citealt{elmegreen78,voit88,deyoung89,mcnamara93}).

The  optical  IFU
spectroscopy of  \citet{oonk10} may lend evidence in  support of this,
as  it  shows  high   velocity  dispersion  molecular  gas  along  the
``sweep-up''  trajectory leading  to  the northern  FUV arc.   Further
evidence for gas displacement by (or perhaps dynamical confinement of)
the  radio   source  can   be  seen  in   the  southern   lobe,  which
anti-correlates  with the  local FUV  distribution. It  is  also worth
noting  that  the ``U''  shaped  loop  of  filamentary star  formation
directly southwest  of the southern  radio lobe is extended  along the
general axis of  the 330 MHz radio source and  the large western X-ray
cavity.   The compelling  morphological  correspondence has  motivated
several past  investigations of jet-triggered star  formation in A2597
\citep{deyoung95,koekemoer99,odea04}.  Each of  these works  found the
model energetically feasible on timescales that make sense in relation
to the radio source \citep{deyoung95}.

It is  worth pursuing these timescale estimates  further, by comparing
the estimated radio source and  X-ray cavity age limits with the young
stellar component dating presented  by \citet{oonk11}, using data from
{\it  HST} and  the  VLT  FORS imager.   That  work compared  measured
FUV/$U$-band ratios with those predicted from \citet{bruzual03} simple
stellar  population  models.   We   show  the  YSP  ages  inferred  by
\citet{oonk11} in the  rightmost panel of Fig.~\ref{fig:agemap}.  Note
that these estimated ages are better considered in the relative rather
than  absolute sense.  The ages  are unavoidably  uncertain  given the
age-metallicity-extinction  degeneracy associated with  measuring ages
of  a young  stellar  component. The  FUV/$U$-band  color is  strongly
dependent upon  the nature and patchiness of  the intrinsic reddening,
which  cannot  be  quantified  on  these  scales.  Furthermore,  while
\ion{C}{iv}   $\lambda$1549  \AA\   resonant line emission  has   so  far   not  been
convincingly  detected in  A2597 \citep{odea04,oonk11},  a significant
contribution  from this  line would  result in  an  underestimation of
stellar ages from the  FUV/$U$-band color. 
We note that \ion{C}{iv} has recently been detected in the {\it HST} Cosmic 
Origins Spectrograph (COS) FUV spectroscopy of M87 \citep{sparks12}.

With these caveats in mind, the  age map shows younger stars nearer to
the  radio  source.  Looking  on  smaller scales,  there  is  possible
evidence of even younger (5 Myr)  stars found (a) at the northern edge
of the  northern radio lobe (where jet-triggered  star formation could
have recently occurred), and (b)  along the projected axis of the host
galaxy stellar isophotes.  When we compare X-ray cavity ages from this
paper with  these YSP ages,  we find that  the range of  young stellar
ages entirely  encompasses the inferred  age range of the  X-ray cavity
network. \citet{oonk11}  found ages for  the YSP ranging  from $5-700$
Myr  old, while our  estimates for  the X-ray  cavity ages  range from
$10-200$  Myr.  The stellar ages also encompass the age 
range estimated for the radio sources (see Section \ref{section:timescalebudget}). 
The stellar ages inferred by \citet{oonk11} are consistent with 
similar estimates by \citet{koekemoer99}.

Two independent studies infer a  young star age range with limits that
are  older  and younger  than  the age  limits  for  the X-ray  cavity
network.  While  acknowledging the many caveats  associated with these
estimates, this could imply that  low levels of star formation ($2-12$
\Msol\   yr\mone)  have  managed   to  persist   even  amid   the  AGN
feedback-driven  excavation of  the X-ray  cavities.  The  stellar age
spatial distribution may further  suggest that star formation was more
spatially  extended in  the past.  This interpretation  is ambiguously
dependent upon the likely fact  that most star formation occurs in the
filaments,  which are  more concentrated  in  the center  than in  the
periphery. Nevertheless, we cannot  rule out the possibility that even
if  the star  formation has  been  persistent during  the current  AGN
feedback episode,  it may be  qualitatively changing with  time (i.e.,
declining and becoming more spatially concentrated).

\section{Concluding Discussion}

The  results presented  in this  paper, considered  in the  context of
three  decades  of previous  work  on  A2597,  motivate the  following
general conclusion: while radio-mode  AGN feedback has injected enough
energy into  the hot  ICM to inhibit  the classical cooling  flow, the
source harbors a  residual cooling flow at 4\% to  8\% of the expected
classical rates,  giving rise to  star formation amid the  $\sim 10^9$
\Msol\ cold molecular  gas reservoir in the nucleus.   While we cannot
rule out  cold gas contributions  from mergers or tidal  stripping, we
suggest that the cooling ICM is likely the dominant supply channel for
cold gas to the nucleus in A2597.

A simple test  of this model can be made  by considering the predicted
residual cooling flow  mass deposition rate with the  observed mass of
the cold  molecular gas reservoir,  along with the star  formation and
black  hole accretion  rates. Together,  these can  be  considered the
ultimate mass  sinks of the residual  cooling flow, so  their mass and
energy budgets should  be consistent with one another.   In Table 3 we
compile  the various  mass, mass  flux, energy,  and  timescale limits
assembled throughout the course of  this analysis. We separate out the
important quantities  related to  inhibition of the  classical cooling
flow (i.e.,  the classical  mass deposition rate  and the  lower limit
kinetic  energy input estimate  based on  X-ray cavity  analysis).  We
then  list the  multiphase constraints  on the  residual  cooling flow
model.   The  predicted residual  mass  deposition  rates are  roughly
$20-40$  \Msol\  yr\mone\  within  30  kpc.  A  steady  residual  mass
deposition rate of this  magnitude would accumulate the observed $\sim
10^{9}$  \Msol\  of gas  in  the central  30  kpc  within $\lae  10^8$
yr.  Taking  the ratio  of  the observed  molecular  gas  mass to  the
observed star formation rate yields  a gas depletion timescale that is
also on the order of 10$^8$ yr.  It is therefore possible for there to
be an approximately steady-state distribution of multiphase gas masses
over the AGN  lifetime, with a possible slow  accumulation of cold gas
at  a  rate  comparable  to  the difference  between  the  local  mass
deposition and star formation rates.

While the proposed model appears  successful in this regard, the above
argument is of course very qualitative.  \citet{gaspari11} presented a
follow  up study  to the  \citet{sharma11} results,  predicting (among
other  things) the  mass budgets  expected from  various  cooling flow
recipes, over time,  in the innermost 20 kpc of  a BCG.  Their results
for an uninhibited cooling flow expectedly overpredict the mass of the
warm  and cold  phases  by  several orders  of  magnitude.  When  they
include AGN  heating at varying efficiencies, their  results fall more
in  line ---  albeit on  the high  end ---  with what  is  observed in
A2597. Their feedback-inhibited residual cooling flow models predict a
roughly  steady  accretion  rate  of  $\sim  10$  \Msol\  yr\mone\  of
$T<5\times10^5$ K  gas within  the central 20  kpc. After 1  Gyr, this
accretion rate  would accumulate  a $\sim 5\times10^{10}$  \Msol\ cold
gas reservoir, which  is an order of magnitude  more massive than what
has  been measured  in  A2597.  However,  as \citet{gaspari11}  notes,
their simulations  do not include  star formation, which  could reduce
the  accumulation rate  of  cold gas  by  an order  of magnitude.   On
roughly the  same scale, the highest  measured SFR for  A2597 is $\sim
10$  \Msol\ yr\mone,  which is  roughly  the cold  gas accretion  rate
predicted on the same scale  by \citet{gaspari11}.  It is also similar
to the  X-ray derived  mass deposition rate  for the  region cospatial
with  the   filaments  of  ionized   gas,  as  discussed   in  Section
\ref{section:filaments}.

\begin{comment}
We show this emission in Fig.~\ref{fig:fuvlya}{\it a}, using the {\it HST}/STIS MAMA observation from \citet{odea04}. We have overlaid 0.5-7 keV adaptively smoothed X-ray contours, as well as 1.3 GHz and  8.4 GHz radio contours in green, blue, and white, respectively.
FUV continuum emission associated with the young stellar component is shown in 
Fig.~\ref{fig:fuvlya}{\it b}, using the  {\it HST}/ACS SBC observation of \citet{oonk10}. 
As both panels are on the same scale, it is apparent that the young stellar 
component is much more spatially compact than the extended Ly$\alpha$ filaments.
We will discuss the implications of this in the next section. 
\end{comment}

\begin{table*}
\begin{minipage}{156mm}
\centering
\caption{A summary of the various mass, mass flux, energy, and timescale limits compiled as part of this paper. 
All values are measured or estimated at or within a central radius of 30 kpc.
(1) Temperature range of the ISM phase (if applicable);
(2) qualitative description of that phase or feature
(3) mass associated with the phase or feature;
(4) power associated with the phase or feature;
(5) estimated mass conversion flux or deposition rate; 
(6) A rough timescale associated with the listed phase or feature. $t_\mathrm{cool}$ is a cooling time associated with the listed cooling flow. $t_\mathrm{heat}$ is a crude estimate of the possible timescale over which X-ray cavity enthalpy could be dissipated as heat in the ISM, limited by the cavity lifetime.  $t_\mathrm{deplete}$ is the cold gas depletion timescale, set by the ratio of the observed cold molecular gas mass and the star formation rate. $t_\mathrm{accum}$ is the time it would take the predicted residual cooling flow mass deposition rate to accumulate the observed cold molecular gas mass. Note that the corresponding timescales ($t_\mathrm{cool}$ vs. ~$t_\mathrm{heat}$ and $t_\mathrm{accum}$ vs.~$t_\mathrm{deplete}$) roughly balance one another.   }
\begin{tabular}{cccccc}
\hline
ISM Phase &
&
$M_\mathrm{phase}\left(R \lae 30~\mathrm{kpc}\right)$ & 
$L\left(R\lae 30~\mathrm{kpc}\right)$ &
$\dot{M}\left(R\lae 30~\mathrm{kpc}\right)$ &
$t$ \\
(K) &
Feature &
(\Msol) &
(ergs sec\mone) &
(\Msol\ yr\mone) &
($\times 10^7$ yr) \\
(1) & (2) & (3) & (4) & (5) & (6)  \\
\hline
\hline
  \multicolumn{6}{c}{{\sc Limits on a Classical Cooling Flow}}\\
\hline
$10^7 - 10^8$    & Classical Cooling flow         &    \nodata     &   $ \left(1-4\right)\times 10^{44} $   &     $\dot{M}_{\mathrm{cool}}\sim 100-500 $                          &  $t_{\mathrm{cool}}\sim 30$    \\
       \nodata           &   AGN Feedback Input  &      \nodata             &       $>1.89\times 10^{44} $              & \nodata  &  $t_\mathrm{heat}\sim 1-40$         \\
\hline
  \multicolumn{6}{c}{{\sc Limits on an Residual Cooling Flow}}\\
\hline       
$10^7 - 10^8$    & Residual Cooling flow         &    \nodata      &  $\left(0.4-3.2\right)\times 10^{43}$    &     $\dot{M}_{\mathrm{cool}}\sim 20-40 $                          &  $t_{\mathrm{accum}}\sim 10$      \\
$10^6$        & FUV Cooling flow               &       \nodata            &        $ \gae 4\times10^{40} $        &                 $\dot{M}_{\mathrm{cool}}\lae 40 $                  &   \nodata          \\
$10^4-10^5$      & Warm Ionized gas        &          $\gae 9.7\pm 0.3\times10^6$   &   $\gae 3.5\times10^{41}$                              &   $\dot{M}_{\mathrm{SFR}}\sim 2-12$   &    \nodata   \\
$10-10^3$           &   Warm/cold molecular gas  &     $\gae 1.8\pm 0.3 \times10^9$     &   $\gae 5\times10^{41}$      &  \nodata   &   $t_\mathrm{deplete}\sim10$       \cr
\hline
\end{tabular}
\end{minipage}
\label{tab:budget}
\end{table*}

\section{Summary}

\label{section:conclusions}

We have presented  a multiwavelength study of the central brightest
cluster galaxy in the cool core cluster Abell 2597. 
The main results of this paper can be summarized as follows. 

\begin{itemize}

\item New {\it Chandra} observations reveal the X-ray 
cavity network to be more extensive than previously known, 
and associated with enough enthalpy to locally inhibit the classical 
cooling flow. 

\item A comparison of estimated cavity and radio source ages
suggests that the AGN duty cycle is near to 100\%, requiring 
a near-steady sub-Eddington flow of gas to the nucleus. 

\item The {\it Herschel}-derived warm dust temperature and mass is 
estimated to be $T_\mathrm{warm~dust}       =      47\pm      1.4$       K      and
$M_\mathrm{warm~dust} = \left(1.7\pm0.6\right)   \times  10^5$  \Msol,
respectively.  The temperature and mass  of the cold dust component is
found     to    be     $T_\mathrm{cold~dust}=20\pm1.7$     K    and
$M_\mathrm{cold~dust} = \left(1.3\pm0.5\right)\times10^7$        \Msol,
respectively. We discuss important uncertainties and assumptions associated with these 
values in Section 4.3.

\item We present an updated CO-inferred cold molecular gas mass of $1.8\pm0.3\times10^{9}$ \Msol, using 
previously unpublished CO(2-1) IRAM 30 m observations. 

\item The newly measured gas and dust masses yield a relatively low, Galactic-type gas-to-dust ratio 
of $\gae 140$, so the gas is dusty.  We argue that mass loss from evolved stars is the most likely source of this substantial dust component. A shielding mechanism is likely required to protect the grains from interaction
with the ambient hot ISM.

\item The warm and cold gas phases are approximately cospatial with the coolest regions 
of X-ray gas. 

\item Dynamical constraints on the central cold gas component do not permit large scale 
rotation or major asymmetric velocity structures, consistent with a scenario wherein the cold 
gas is accreted with low net angular momentum, inconsistent with what is expected from a merger.

\item The theoretical thermal instability threshold lies just outside the observed $\lae 30$ kpc X-ray cavity network. Molecular and ionized, star forming gas lies interior to both. This result is consistent 
with theoretical predictions of the entropy threshold at which ICM cooling begins to form thermally unstable cold clouds and filaments.

\item X-ray derived mass deposition rates along regions cospatial with star forming filaments 
are consistent with the locally estimated star formation rates, strongly suggesting a causal connection. 

\item The young stellar  component  occupies an  age range  that is
apparently wider than that for  the X-ray cavities. This could suggest
that star formation has  persisted amid the feedback-driven excavation
of the  X-ray cavity  network. Localized sites  of star  formation may
also have been  triggered by the propagating radio  source. 

\end{itemize}

We conclude that  a residual 
cooling flow with a strength of   4\% to 8\% of the expected classical mass deposition rates
is the dominant contributor 
of the cold gas reservoir fueling star formation and AGN activity in the Abell 2597 BCG.

\section*{Acknowledgments} The authors thank Drs. Elaine Sadler, Robert Laing, Andy Robinson, 
Joel Kastner, and Bill Sparks for thoughtful discussions. 
We also thank the anonymous referee for constructive feedback. 
G.~R.~T.~is grateful  to R.~A.~S.,  and acknowledges support  from the
NASA/NY  Space  Grant  Consortium,  as  well as  a  European  Southern
Observatory  (ESO)   Fellowship  partially  funded   by  the  European
Community's Seventh Framework  Programme (/FP7/2007-2013/) under grant
agreement No.~229517.  Partial support was provided by NASA through an
award issued  by JPL/Caltech, as  well as the Radcliffe  Institute for
Advanced   Study  at   Harvard  University.    T.~E.~C.~was  partially
supported by NASA through {\it  Chandra} award G06-7115B issued by the
{\it Chandra} X-ray Observatory Center for and on behalf of NASA under
contract NAS8-39073. Basic research  into radio astronomy at the Naval
Research  Laboratory is  supported  by 6.1  Base funds.   C.~L.~S.~was
supported  in part  by NASA  {\it  Herschel} Grants  RSA 1373266,  RSA
P12-78175  and {\it Chandra}  Grant G01-12169X. A.~C.~F.~thanks the Royal Society.   
This paper  is based
upon observations  with the {\it Chandra X-ray  Observatory}, which is
operated  by  the Smithsonian  Astrophysical  Observatory  for and  on
behalf  of NASA  under
contract NAS8-03060.  We also  present observations made with the {\it
  Herschel  Space Observatory},  a European  Space  Agency Cornerstone
Mission with significant  participation by NASA.  We make use of
previously published  observations by  the NASA/ESA {\it  Hubble Space
  Telescope}, obtained at the Space Telescope Science Institute, which
is  operated  by  the  Association  of Universities  for  Research  in
Astronomy, Inc., under NASA  contract 5-26555.  
The National Radio Astronomy Observatory is a facility of the National Science 
Foundation operated under cooperative agreement by Associated Universities, Inc.
We have made extensive
use of  the NASA Astrophysics  Data System bibliographic  services and
the NASA/IPAC  Extragalactic Database, operated by  the Jet Propulsion
Laboratory,  California Institute of  Technology, under  contract with
NASA.

\bibliographystyle{mn2e}
\bibliography{a2597_complete}

\label{lastpage}

\end{document}